\documentclass{amsproc}
\usepackage[geometry]{ifsym}

\newtheorem{theorem}{Theorem}[section]
\newtheorem{lemma}[theorem]{Lemma}

\theoremstyle{definition}

\theoremstyle{remark}

\newcommand{\ds}{\displaystyle}

\numberwithin{equation}{section}
\DeclareMathOperator*{\Max}{max}
\DeclareMathOperator*{\argmax}{argmax}
\begin{document}

\title{Efficient Auctions With Common Values}

%    Remove any unused author tags.

%    author one information
\author{Andrei Laurentiu Ciupan}

\subjclass[2000]{Primary }
%    For articles to be published after 1 January 2010, you may use
%    the following version:
%\subjclass[2010]{Primary }

\keywords{}

\date{}

\begin{abstract}

Consider the problem of allocating goods to buyers through an auction. An auction is efficient if the resulting allocation maximizes total welfare, conditional on the information available. If buyers have private values, the Vickrey-Groves-Clarke mechanism is efficient. If buyers have common values and a buyer's information can be summarized as a one-dimensional signal, Dasgupta and Maskin present an efficient auction. We construct an efficient auction mechanism in case buyer information is multidimensional, for a restricted class of valuation functions, and we prove which of the assumptions made are necessary for the existence of an efficient mechanism.
\end{abstract}

\maketitle

\section{Introduction}

Assume there is a set $\ds N =\{1, 2, \dots n\}$ of $\ds n$ buyers, and a set $\ds M =\{1,2, \dots m\}$ of $\ds m$ goods. Each buyer $\ds i$ observes $\ds m$ one-dimensional signals $\ds s_{i1}, s_{i2}, \dots s_{im}$, where $\ds s_{ik}$ is buyer $\ds i$'s signal corresponding to good $\ds k$. For every $\ds i\in N, k\in M$, let $\ds \mathcal S_{ik}\subseteq\mathbb R$ denote the space of all possible signals that buyer $\ds i$ can observe for good $\ds k$. Further, for every $\ds k\in M$, let $\ds \mathcal S_k =\times_{i\in N} \mathcal S_{ik}$. Let $\ds T_i = \times_{k\in M} \mathcal S_{ik}$ denote the space of all signals that buyer $i$ can observe for all goods. For any buyer $\ds i\in N$ and any good $\ds k\in M$, let $\ds \mathcal T_{i, k}$ be the space of signals relevant to buyer $\ds i$ 's valuation of good $\ds k$, and let $\ds \mathcal T_i =\times_{k\in M} \mathcal T_{i, k}$ the set of all signals relevant to buyer $\ds i$'s valuation of all goods. Finally, let $\ds v_{i,k} : \mathcal T_{i, k}\mapsto \mathbb R$ be buyer $\ds i$'s valuation function of good $\ds k$.

We may consider valuations of more than one good from the set $\ds M$. For a buyer $\ds i$ and a subset of goods $\ds S\subseteq M$, let $\ds \mathcal T_{i, S} = \times_{k\in S} \mathcal T_{i, k}$ be the space of signals relevant to buyer $\ds i$'s valuation of the set $\ds S$, and let $\ds v_{i,S}: \mathcal T_{i, S}\mapsto \mathbb R$ be buyer $\ds i$'s valuation function for the goods in the set $\ds S$. The valuation of an empty set $\ds v_{i, \emptyset} $ is null for every buyer $\ds i\in N$.

If buyers have private values, then $\ds \mathcal T_{i, k} = \mathcal S_{ik}$, for all $\ds i\in N$ and $\ds k\in M$.

If a buyer's valuation of a given good depends on the other buyers' signals, i.e. if buyers have common values, then $\ds \mathcal T_{i, k} = \mathcal S_k$, for all  $\ds i\in N$ and $\ds k\in M$.

Let us illustrate common and private values with two examples.

\emph{Example 1.} Consider two investors competing for an apartment complex. Each investor $i$ observes the current value of the apartment complex, denoted $\ds s_i$, has a projected return on investment $\ds r_i$ and a faces a fixed cost $\ds c_i$. Therefore, buyer $\ds i$'s valuation of the auctioned good is 

\begin{equation}
\ds v_i(s_i) = (1+r_i)\cdot s_i - c_i
\end{equation}

Since investor $\ds i$'s valuation is independent of any other investor's, this is a case of private values.

\emph{Example 2.} Consider two amateur collectors competing for two Van Gogh paintings, labeled $\ds A$ and $\ds B$. Each collector is only interested in buying only one of the two paintings and has ordered a private evaluation of each painting, with buyer $\ds i$ obtaining an estimated valuation of $\ds s_{iA}$ for good $\ds A$ and $\ds s_{iB}$ for good $\ds B$. However, each collector considers that his or her valuation is likely incomplete, and considers the other collector's estimations when deciding how much each painting is worth, taking a linear combination of the $2$ private estimated evaluations. The collector valuations can be expressed as
\begin{equation}
\ds v_{1A}(s_{1A}, s_{2A}) = s_{1A} +\frac{1}{2}s_{2A},\mbox{   } v_{1B}(s_{1B}, s_{2B}) = s_{1B} +\frac{1}{2}s_{2B}
\end{equation}
\begin{equation}
\ds v_{2A}(s_{1A}, s_{2A}) = s_{2A} +\frac{1}{3}s_{1A},\mbox{   }  v_{2B}(s_{1B}, s_{2B}) = s_{2B} +\frac{1}{3}s_{1B}
\end{equation}

This is an example of common values, since a buyer's valuation is affected by the other buyer's signal. 

Note that it seems reasonable to assume that the full information determining a buyer's valuation for a set of goods is not a one-dimensional value but a multidimensional vector, with one signal component for each good. Consider the example above: if the goods are different, then the full valuation information is a function of $\ds s_{1A}$ for good $\ds A$ and $\ds s_{1B}$ for good $\ds B$, and these two signals need not be related.

Even though the multidimensional signals assumption is more general than a one-dimensional signal assumption, the authors in \cite{maskin00} show that we cannot have, in general, an efficient auction with multidimensional signals. We deal with the necessary assumptions in section $\ds 3$.

Throughout this paper we assume that signal values are private information, i.e. each buyer only knows his or her signal values and has no information on the other buyers' signal values. The auction designer also has no knowledge about buyers' signal values.

An \emph{allocation} is a partition $\ds M =\sqcup_{i\in N} S_i$, such that for all $\ds i\in N$, buyer $\ds i$ receives the goods in the set $\ds S_i$. The welfare of an allocation $\ds M =\sqcup_{i\in N} S_i$ is defined as the total sum of valuations, given the buyers' signals:
\begin{equation}
\ds W(S_1, S_2, \dots S_n) =\sum_{i\in N} v_{i, S_i} (t_{i, S_i}),
\end{equation} where $\ds t_{i, S_i}$ is the relevant set of signals for buyer $\ds i$'s valuation of the goods in the set $\ds S_i$, as derived from the overall set of signals $\ds \{ s_{ik} | i\in N, k\in M\}$. 

Recall that for each buyer $\ds i\in N$, $\ds T_i =\times_{k\in M} \mathcal S_{ik}$ is the space of all signals that buyer $\ds i$ can observe for all goods. For a given set of signals $\ds t_i\in T_i$, let $\ds \mathcal \sigma_i(t_i)$ define buyer $\ds i$'s strategy in this case, and let $\ds \Sigma_i $ denote the set of all possible strategies that buyer $\ds i$ could have. Note that a strategy $\ds \sigma_i\in \Sigma_i$ need not be a real-valued function. It is simply a defined set of actions that buyer $\ds i$ will undertake for any given set of signals $\ds t_i \in T_i$. 

The auction designer interacts with buyers according to their strategies $\ds \sigma_j(t_j) $ for all buyers $\ds j
\in N $ and decides how to allocate the goods being auctioned and how much to charge each buyer. If as a result of an auction buyer $\ds i$ receives a set $\ds S$ of goods with valuation $\ds v_i(S)$ and is required to make a payment $\ds P_i(S)$, his or her utility is simply the valuation of the goods received minus the payment, 

\begin{equation}
\ds U_i(S, P_i(S)) = v_i(S) - P_i(S)
\end{equation} For a given set of buyer signals $\ds t_1, t_2, \dots, t_n$ and a given set of buyer strategies $\ds \sigma_1, \sigma_2, \dots, \sigma_n$, let 

\begin{equation}
U_i(\sigma_1, \sigma_2, \dots, \sigma_n, t_1, t_2, \dots ,t_n) 
\end{equation}

denote buyer $\ds i$'s expected utility as a result of the auction allocation and payment rules, following $\ds (1.5)$. 

We say that an $\ds n$-tuple of strategies $\ds \sigma^* = (\sigma_1^*, \sigma_2^*, \dots, \sigma_n^*)\in\Sigma_1\times\Sigma_2\cdots\times \Sigma_n$ constitutes a Nash Equilibrium if for all buyers $\ds i\in N$ and for all $\ds (t_1, t_2, \dots t_n) \in T_1\times  T_2\cdots\times T_n$, we have

\begin{equation}
U_i(\sigma^*, t_1, t_2, \dots ,t_n) \geq U_i(\sigma_1^*, \sigma_2^*, \dots,\sigma_{i-1}^*, \sigma_i, \sigma_{i+1}^*, \dots,  \sigma_n^*, t_1, t_2, \dots ,t_n), \mbox{for all } \sigma_i\in \Sigma_i
\end{equation}

This means that given the other buyers' strategies $\ds \sigma_{-i}^*$, the strategy $\ds \sigma_i^*$ is the one with the highest utility for buyer $\ds i$ for all possible signal values $\ds (t_1, t_2, \dots t_n) $ that the buyers could have.  In this paper we only deal with \emph{pure strategies}, i.e. buyer strategies which are not randomized.

An auction is \emph{efficient} if there is a Nash Equilibrium of the buyers' bidding strategies which leads to the welfare-maximizing allocation. We continue by describing efficient auction mechanisms for private and common values. 

We first deal with the case of private values, then overview the Dasgupta-Maskin \cite{maskin00} results and move forward to the case of multidimensional signals and common values. Finally, in case valuation functions are linear and each buyer gains no additional benefit from having more than one good, we present an efficient auction mechanism.

\section{Private Values}

In the case of private values, as in Example 1, the Vickrey- Clarke-Groves mechanism is efficient. We describe the mechanism and prove that it is indeed efficient.

Assume we have a set of $\ds N$ buyers and $\ds M$ goods to be auctioned. The mechamism is as follows:

\begin{enumerate}

\item \emph{Bidding.} Each buyer $\ds i\in N$ selects a subset $\ds X_i \subseteq 2^M$ and submits bids $\ds b_{i} ( S)$ for each $\ds S\in X_i$. For all subsets of $\ds S'\subseteq M$ for which buyer $\ds i$ did not submit a bid, it is considered that $\ds b_{i}( S') =0$.

\item \emph{Allocation.} Given buyers' bids, consider the allocation $\ds S = (S_1, S_2, \dots, S_n)$ which maximizes welfare as defined in $\ds (1.4)$, under the hypothesis that the bids which the buyers submit represent their actual valuations for each set of goods. If there are more such allocations, uniformly select one of them at random. Each buyer $\ds i$ receives the goods in the set $\ds S_i$.

\item \emph{Payment.} Let $\ds S'_{-i} =  (S_1(i), S_2(i), \dots, S_n(i)) $ be the welfare-maximizing allocation of the $\ds M$ goods to the buyers $\ds N - \{i\}$, under the same hypothesis from step $\ds (2)$ above, and let $\ds W(S_{-i}') = \sum_{j\in N, j\neq i} b_{j}( S_j(i))$ be the apparent welfare from this allocation. Note that $\ds S_i(i) =\emptyset$. If $\ds  S = (S_1, S_2, \dots, S_n)$ is the selected allocation from step $\ds (2)$, with apparent welfare $\ds W(S) = \sum_{j\in N} b_{j}( S_j)$, then buyer $\ds i$ pays $\ds P_i(S) = W(S_{-i}') - W(S) + b_i(S_i)$.
 \end{enumerate}
As in all the auction mechanisms that we present in the paper, we first prove that the allocation and payment mechanisms are well-defined, then prove that truthful bidding is a Nash Equilibrium. Let's first prove that the payment mechanism is well-defined.

\begin{lemma}
For any buyer $\ds i \in N$, the payment $\ds P_i(S)$ is nonnegative and does not depent on the bid that he or she makes. It only depends on the selected allocation in step (2) .
\end{lemma}

\emph{Proof.} For any buyer $\ds j\in N$ and any set $\ds S\subseteq M$, let $\ds v_j(S)$ denote buyer $\ds j$'s valuation of the set of goods in $\ds S$. Let $\ds S_i$ and $\ds S_{-i}'$ be the allocations selected at steps $\ds (2)$ and $\ds (3)$ from the mechanism above. Note that $\ds  P_i(S) = W(S_{-i}') - W(S) + b_i(S_i) = \sum_{j\in N, j\neq i} \left(b_j(S_j(i)) - b_{j}(S_j)\right) $, so indeed the payment does not depend on $\ds b_{i}(S_i)$. Furthermore, since $\ds S_{-i}'$ is the welfare-maximizing allocation of the goods in $\ds M$ to the buyers in $\ds N - \{i\}$, we must have $\ds  \sum_{j\in N, j\neq i} b_j(S_j(i)) \geq  \sum_{j\in N, j\neq i} b_{j}(S_j)$, so indeed the payment is nonnegative, as desired $\ds\square$.

We will prove that this auction is efficient.

\begin{theorem} \cite{vickrey61, groves73,clarke71}Truthful bidding, i.e. bidding one's true valuation of every subset of the goods in $\ds M$, represents a Nash Equilibrium.

\end{theorem}
\emph{Proof.} Fix a buyer $\ds i\in N$, and assume that all other buyers bid truthfully, i.e $\ds b_j(S) = v_j(S)$ for all buyers $\ds j\neq i$ and all subsets $\ds S\subseteq M$. Let $\ds S = (S_1, S_2, \dots S_n)$ be the allocation of the goods according to the auction above if all buyers bid truthfully, with $\ds S_i$ be the set of goods that buyer $\ds i$ obtains (with the observation that $\ds S_i$ can be empty). In this case, buyer $\ds i$ makes a payment of
$\ds P_i(S) = W(S_{-i}') - W(S) + b_i(S_i)$, where $\ds S'_{-i} =  (S_1(i), S_2(i), \dots, S_n(i)) $ is the welfare-maximizing allocation of the $\ds M$ goods to the buyers $\ds N - \{i\}$.

Note that buyer $\ds i$'s utility from truthful bidding is $\ds v_{i} (S_i) - P_i(S) = v_i(S_i) + W(S) - W(S_{-i}') - v_i(S_i)$, which can be rewritten as 
\begin{equation}
\ds W(S) - W(S_{-i}')
\end{equation} Since $\ds S$ is the welfare-maximizing allocation, buyer $\ds i$'s utility is nonnegative in this case, and therefore he or she has no incentive to bid so that he or she doesn't receive any goods.

Now assume that buyer $\ds i$ bids so that the welfare-maximizing allocation of the goods in $\ds M$ to the buyers in $\ds N$, as calculated in step $\ds (2)$ of the auction above, is $\ds S' = (S'_1, S'_2, \dots , S_n')$. Note that since all other buyers except for $\ds i$ bid truthfully, so $\ds S_{-i}'$, the welfare-maximizing allocation of the goods in $\ds M$ to buyers in $\ds N -\{i\}$, is unchanged. Buyer $\ds i$ pays $\ds P_i(S') =  W(S_{-i}') - W(S') + b_i(S'_i)$, and his utility in this case is 
\begin{equation}
\ds v_i(S'_i)  -  W(S'_{-i}) + \sum_{j\in N, j\neq i} v_j(S'_j)
\end{equation} Let us compare this to the utility obtained from truthful bidding, $\ds W(S) - W(S_{-i}')$. Since $\ds S$ is the welfare-maximizing allocation under truthful bidding, we must have $\ds W(S) \geq  v_i(S'_i) + \sum_{j\in N, j\neq i} v_j(S'_j)$, which indeed implies that the utility obtained from truthful bidding is greater than or equal to the utility obtained from any other bidding. Thus truthful bidding represents a Nash Equilibrium $\ds\square$.

Note that a very similar approach allows us to prove that truthful bidding is in fact weakly dominant, i.e. regardless of the other buyers' strategies, truthful bidding is always a best response for a given buyer.

We will see that there is a common characteristic and intuition behind all efficient auctions: have buyers submit their signals, valuations, or equivalent messages which would allow valuations to be calculated, allocate goods according to the apparent welfare-maximizing allocation, and select a payment which induces truthful bidding in Nash Equilibrium. 

It is the payment mechanism that is the most important in efficient auctions. The idea behind the Vickrey-Groves-Clarke payment mechanism is that each buyer $\ds i$ has to pay the marginal effect of his or her presence in the auction to the other buyers: if buyer $\ds i$ were not present in the auction, the buyers' welfare is equal to $\ds W_{-i}$. With buyer $\ds i$ present in the auction, denote total welfare $\ds W $, and let $\ds v_i(S_i)$ be buyer $\ds i$'s utility. Therefore, with buyer $\ds i$ present in the auction, the total welfare that the $\ds N-\{i\}$ buyers receive decreases to $\ds W - v_i(S_i)$, and the marginal effect of buyer $\ds i$'s presence in the auction is $\ds W_{-i} - W + v_i(S_i)$. 

This is the payment that we want to induce to buyer $\ds i$. Note that this payment allows buyer $\ds i$'s utility (valuation minus payment) to be a function of total welfare, as seen in relations $\ds (2.1)$ and $\ds (2.2)$. Therefore the payment mechanism forces each buyer's attempt to maximize his or her own utility to be equivalent to maximizing overall welfare, leading the mechanism to be efficient, as proved above. We will see variations of these ideas throughout the next auction designs.

Let us now deal with the case of common values.

\section{Common values and one-dimensional signals}

In \cite{maskin00}, Dasgupta and Maskin deal with the case of common values and one-dimensional signals for the general case of any number of buyers and any number of goods. In this section we present the case of $\ds n\geq 2$ buyers and one good. Assume that buyer $\ds i$ receives signal $\ds s_i$, and that his or her valuation is a function of all other buyers' signals, $\ds v_i(s_1, s_2, \dots s_n) $. For simplicity let $\ds \mathcal S_i =\mathbb R$, i.e. assume that the set of all possible signals for each buyer is the real line. The authors make the following two assumptions on valuation functions: each buyer $\ds i$'s valuation must be increasing in $\ds s_i$, and the marginal effect of signal $\ds s_i$ on buyer $\ds i$'s valuation is larger than the marginal effect of signal $\ds s_i$ on any other buyer's valuation. These can be expressed as 

\begin{equation}
\frac{\partial v_i}{\partial s_i} (s_1, s_2, \dots, s_n) >0, \mbox {for all } (s_1, s_2, \dots, s_n)\in \mathbb R^n, \mbox{ and}
\end{equation}

\begin{equation}
\ds \mbox{ For all } i, j \in N, i\neq j, \mbox{ we have } \frac{\partial v_i}{\partial s_i} (s_1, s_2, \dots, s_n) >  \frac{\partial v_j}{\partial s_i} (s_1, s_2, \dots, s_n) 
\end{equation}

at any point where $\ds v_i(s_1, s_2,\dots, s_n) = v_j(s_1, s_2, \dots, s_n) = \Max\limits_{k\in N} v_k(s_1, s_2, \dots s_n) $

Further, assume that valuation functions are common knowledge among the buyers, i.e. buyer $\ds i$ knows buyer $\ds j$'s valuation function for any $\ds i, j \in N$. The auction designer need not know the functional forms. We are looking for an auction design which would determine truthful bidding and allow buyer valuations to be calculated. The following mechanism is proposed for two buyers:

\begin{enumerate}

\item \emph{Bidding.} Each buyer $\ds i$ submits a bid function $\ds b_i :\mathbb R\mapsto \mathbb R$, such that 
\begin{equation}
\ds |b_i'(x) | <1 \mbox{ for all } x\in\mathbb R.
\end{equation}

\item \emph{Allocation.} Given the bid functions above, select the fixed point $\ds (v_1^{\circ}, v_2^{\circ})$ such that 
\begin{equation}
\ds (v_1^{\circ}, v_2^{\circ}) = (b_1(v_2^{\circ}), b_2(v_1^{\circ}))
\end{equation}

If there is no fixed point, the auction terminates and the good is not allocated.
Condition $\ds (3.3)$ assures that there is at most one fixed point, since relation $\ds (4.4)$ can be rewritten as $\ds b_1(b_2(v_1^{\circ})) = v_1^{\circ}$ , and the function $\ds b_1\circ b_2 $ also satisfies condition $\ds (3.3)$, implying that it must have at most one fixed point.
If $\ds v_1^{\circ} > v_2^{\circ}$, then buyer $\ds 1$ receives the good. If  $\ds v_1^{\circ} < v_2^{\circ}$, buyer $\ds 2$ receives the good. If $\ds v_1^{\circ} = v_2^{\circ}$, the good is offered uniformly at random to one of the buyers.

\item \emph{Payment.} If buyer i receives the good, he or she makes a payment $\ds v_1^{*}$, where $\ds v_1^{*}$ is a point for which $\ds v_1^{*} = b_2(v_1^*)$
\end{enumerate}

Assume that buyer $\ds 1$ observes signal $\ds s_1$ and buyer $\ds 2$ observes signal $\ds s_2$. In this auction, function $\ds b_1$ of buyer $\ds 1$ represents a truthful bid if and only if

\begin{equation}
\ds b_1(v_2(s_1, s_2)) = v_1(s_1, s_2) \mbox{ for all } s_2\in\mathbb R, 
\end{equation}
with the simmetric definition for buyer $\ds 2$. Note that condition $\ds (3.1)$ assures that relation $\ds (3.5)$ is well-defined.  Note that as a result of truthful bidding, the equilibrium point in $\ds (v_1^{\circ}, v_2^{\circ}) $ is precisely $\ds (v_1(s_1, s_2), v_2(s_1, s_2))$. The authors in \cite{maskin00} prove that truthful bidding is a Nash Equilibrium of this auction, and that as a result of truthful bidding, the good is allocated to the buyer with the highest valuation $\ds v_i(s_1, s_2)$.

The case of $\ds n\geq 3$ buyers, the auction mechanism is in the same vein as above, but there is more care regarding multiple potential fixed points. Buyers are again required to submit bid functions contingent on the other buyers' valuations, but since the valuations don't necessarily uniquely identify one valuation, as in equation $\ds (3.5)$ above, buyers are required to submit bid correspondences, representing the set of valuations they could potentially have if the other buyers have given valuations.

The auction design is as follows:

\begin{enumerate}
\item \emph{Bidding.} Each buyer $\ds i\in N$ submits a bid correspondence $\ds b_i : \mathbb R^{n-1} \longrightarrow \mathbb R$
\item \emph{Allocation.} A fixed point $\ds v^{\circ} = (v_1^{\circ}, v_2^{\circ}, \dots, v_n^{\circ}) $ is chosen so that

\begin{equation}
\ds v_i^{\circ} \in b_i(v_{-i}^{\circ}), \mbox{ for all } i\in N
\end{equation}

If there is no such fixed point, the good is not awarded and no buyer makes a payment. If there are multiple such n-dimensional fixed points $\ds v^{\circ 1} = (v^{\circ 1}_1, v^{\circ 1}_2, \dots , v^{\circ 1}_n) , v^{\circ 2} =  (v^{\circ 2}_1, v^{\circ 2}_2, \dots , v^{\circ 2}_n), \dots,  v^{\circ k} =  (v^{\circ k}_1, v^{\circ k}_2, \dots , v^{\circ k}_n)$, then the bid functions are made public, and each buyer $\ds i$ chooses one of the $\ds k$ values $\ds v^{\circ 1}_i, v^{\circ 2}_i, \dots, v^{\circ k}_i .$  If the buyers do not agree on the same fixed point, the good is not awarded and no buyer makes a payment. If all buyers agree on the same fixed point  $\ds v^{\circ} = (v_1^{\circ}, v_2^{\circ}, \dots, v_n^{\circ}) $, the buyer satisfying $\ds i =\argmax_{j\in N} v_{j}^{\circ}$ receives the good. If there is a tie, the good is allocated uniformly at random among the buyers with the highest $\ds v_i^{\circ}$.

\item \emph{Payment} Let $\ds v^{\circ} = (v_1^{\circ}, v_2^{\circ}, \dots, v_n^{\circ}) $ be the fixed point selected at step 2 above. If the good was not allocated to buyer $\ds i$, he or she makes no payment. If the good was allocated to buyer $\ds i$, he or she makes a payment of 

\begin{equation}
\Max_{j\neq i, j\in N} v_j^*,
\end{equation}
where $\ds v^* = (v_1^*, v_2^*, \dots v_n^*)$ is a vector such that 
\begin{equation}
\ds v_k^*\in b_k(v_{-k}^*) \mbox{ for all } k\neq i, k\in N
\end{equation} and 

\begin{equation}
v_i^* = \Max_{j\in N, j\neq i} v_j^*
\end{equation}

If no such vector $\ds v^*$ exists, the good is not awarded and no buyer makes a payment. If there are multiple such vectors, the bids are made public and each buyer selects one of the possible vectors satisfying $\ds (3.8)$ and $\ds (3.9)$ above. If they all agree on the same vector, buyer $\ds i$ makes a payment according to $\ds (3.7)$. If the buyers do not all agree, the good is not awarded and no buyer makes a payment.

\end{enumerate}

The authors prove that truthful bidding constitutes a Nash Equilibrium, and hence the auction is efficient. Truthful bidding of buyer $\ds i$ with signal $\ds s_i$ is characterized by bidding $\ds b_i (v_{-i}) = \{ v_i \in \mathbb R |\mbox{ there exist signals } s'_{-i} \in \mathbb R^{n-1} \mbox{ such that } v_i(s_i, s'_{-i}) = v_i \mbox{ and } v_j(s_i, s'_{-i}) = v_j \mbox{ for all } j\in N, j\neq i\}$. Moreover, since there might be a chance of multiple fixed points in either step $(2)$ or step $(3)$ in the auction above, those strategies need to be defined as well. Here is how the authors deal with the multiple points of step (2) :

If there is a unique vector $\ds s_{-i}' $ such that, for all buyers $\ds j\neq i$ and all $\ds v_{-j} \in \mathbb R^{n-1} $, the bid functions (which are made public in case there are multiple fixed points in step $\ds 2$) satisfy the property that $\ds b_{j}(v_{-j}) $ represents a truthful bid of buyer $\ds j$ as if his signal were $\ds s_{j}'$, then buyer $\ds i$ chooses the value satisfying $\ds v_i^{\circ} = v_i(s_i, s_{-i}') $. If there are no such vectors $\ds s_{-i}'$ or multiple such vectors, then buyer $\ds i$ chooses one of the fixed points described in step $\ds (2)$ uniformly at random.

Similarly, truthful bidding in case of multiple fixed points at step $\ds 3$ is defined as follows:

If there is a unique vector $\ds s_{-i} '$ such that, for all buyers $\ds j\neq i$ and all $\ds v_{-j} \in\mathbb R^{n-1}$, the bid functions satisfy the property that $\ds b_{j} (v_{-j}) $ represents a truthful bid of buyer $\ds j$ as if his or her signal were $\ds s_j'$, and additionally $\ds v_i(s_i, s'_{-i}) = \Max _{j\in N, j\neq i} v_{j}(s_i, s_{-i}')$, then the buyer selects the fixed point $\ds v_i^* = v_i(s_i, s'{-i})$. If there no such vectors $\ds s'_{-i}$ or multiple such vectors, then buyer $\ds i$ randomizes uniformly between the available fixed points.

In case there is only one auctioned good, the Dasgupta-Maskin mechanisms are efficient. In case there is more than one good to be auctioned, the set of signals relevant for a group of two or more goods will not be one-dimensional anymore. However, under the assumption that there exist real numbers $\ds t_1, t_2, \dots t_n$ such that for any buyer $\ds i$ and any set $\ds S$ of goods, buyer $\ds i$'s valuation of $\ds S$ can be expressed as a function of $\ds t_1, t_2, \dots t_n$,the authors extend the one-good mechanism above to any set of goods.

In the next sections we deal with the case where valuation functions cannot be summarized by a one-dimensional vector for every buyer. In the case where no buyer receives any marginal utility from having more than one good from the set we present an efficient auction mechanism. We can imagine example $\ds (2)$ in the introduction as a suitable case, or consider spectrum auctions, where the government auctions airwave frequency intervals to companies and no company is interested in two or more airwave frequency intervals, since they would simply use just one of the allocated offerings.

\section{Common values and multidimensional signals}

Let us focus on the case where each buyer gets no additional benefit of having more than one good. Formally, this can be represented in the valuation function as:

For every $\ds i\in N$, every $\ds S =\{a_1, a_2, \dots ,a_k\} \subseteq M$ and every $\ds s_{a_t} \in \mathcal T_{i,a_t}$ :

\begin{equation}
\ds v_{i, S} (s_{a_1}, s_{a_2}, \dots s_{a_k}) = \Max\limits_{1\leq t\leq k} v_{i, a_t} (s_{a_t}),
\end{equation}

i.e. every buyer's valuation of a set of goods simply equals the valuation of the highest-valued good in that set.

Throughout this section, assume that $n\geq m$, i.e. there are at least as many buyers as goods. Also assume that the auction designer knows buyer valuations for each good. This assumption will be dropped in the next section.

We assume that buyers have common values and valuations are linear. Moreover, assume that for any buyer $\ds i$, the marginal effect of his or her signal on any other buyer's valuation for any good is independent of the other buyer and of the good. This can be translated as:

For any $\ds i\in N$, there exist linear functions $\ds f_i, w_i : \mathbb R \mapsto \mathbb R$ such that every $\ds w_i$ is strictly increasing and

\begin{equation}
v_{i, K} (s_{1K}, s_{2K},\dots ,s_{nK} ) = w_i(s_{iK}) +\sum_{j\neq i} f_j(s_{jK}), \forall \ds K\in M, \forall (s_{1K}, s_{2K}, \dots s_{nK})\in \mathcal S_K
\end{equation}

We require that the marginal effect of every buyer's signal on his or her valuation is larger than the marginal effect of the same signal on any other buyer's valuation, i.e. for all $\ds K\in M$, for all $\ds i\neq j$ and for all $\ds  (s_{1K}, s_{2K}, \dots s_{nK})\in \mathcal S_K$
\begin{equation}
\ds \frac{\partial v_{i,K}}{\partial s_{iK}}(s_{1K}, s_{2K},\dots s_{nK} ) > \frac{\partial v_{j,K}}{\partial s_{iK}}(s_{1K}, s_{2K},\dots, s_{nK} ),
\end{equation} or equivalently, in light of relation $\ds (4.2)$:

\begin{equation}
w_i'(\cdot) > f_i'(\cdot), \forall i\in N
\end{equation}

Finally, we require that all valuations are nonnegative for a given set of buyer signals $\ds S =\{s_{iK} | i\in N, K\in M\} $.

\begin{equation}
v_{i, k}  (s_{1k}, s_{2k},\dots s_{nk} ) \geq 0, \forall i\in N, \forall k\in M. 
\end{equation}

Relation $ (4.3)$ is simply an extension of assumption $\ds (3.2)$ in the previous section, where for simplicity we assumed that it holds for all signal values $\ds  (s_{1K}, s_{2K}, \dots s_{nK})\in \mathcal S_K$, and not just when we have  $$\ds v_{i, K} (s_{1K}, s_{2K}, \dots ,s_{nK}) = v_{j, K} (s_{1K}, s_{2K}, \dots ,s_{nK}) =\Max\limits_{t\in N} v_{t, K} (s_{1K}, s_{2K}, \dots ,s_{nK}) .$$

Since valuations are linear, the two are equivalent.

Let us show that this assumption is in fact necessary. For simplicity and conciceness, we prove the following weaker result:

\begin{lemma}
Consider $\ds N$ buyers, each with signal space $\ds \mathcal S_i $ an open interval of $\ds \mathbb R$ for one good that is being auctioned, and valuation functions that are increasing in each buyer's signal. If an efficient auction exists, then we must have $$\ds \frac{\partial v_i}{\partial s_i} (s_1, s_2, \dots, s_n) \geq  \frac{\partial v_j}{\partial s_i} (s_1, s_2, \dots, s_n) $$ at any point where  $\ds v_i(s_1, s_2, \dots, s_n) = v_j(s_1, s_2, \dots, s_n) =\Max_{k\in N} v_k(s_1, s_2, \dots, s_n)$
\end{lemma}

\emph{Proof.} Let $\ds (\sigma_1, \sigma_2, \dots s_n) $ be a Nash Equilibrium of this auction such that the good is offered to the buyer with the highest valuation as a result of these strategies, for any signal values $\ds (s_1, s_2,\dots s_n) \in \mathbb \mathcal S_1\times\mathcal S_2\cdots\times\mathcal S_n$. Assume there is a point $\ds (s_1, s_2, \dots, s_n)$ for which $\ds v_i(s_1, s_2, \dots, s_n) = v_j(s_1, s_2, \dots, s_n) =\Max_{k\in N} v_k(s_1, s_2, \dots, s_n)$, and the good is offered to buyer $\ds i$ with positive  probability $\ds p_i$ as a result of strategies $\ds \sigma_1(s_1), \sigma_2(s_2), \dots, \sigma_n(s_n)$. 

Further, let $\ds P_i(\sigma_1(s_1), \sigma_2(s_2), \dots, \sigma_n(s_n) )$ be buyer $\ds i$'s expected payment to the auction designer for strategies $\ds \sigma_1, \sigma_2,\dots, \sigma_n$ and buyer signals $\ds (s_1, s_2, \dots, s_n)$.

If $\ds \frac{\partial v_i}{\partial s_i} (s_1, s_2, \dots, s_n) <  \frac{\partial v_j}{\partial s_i} (s_1, s_2, \dots, s_n) ,$ then there is a point $\ds s_i'\in\mathcal S_i$ such that $\ds s_i' > s_i$ and $$\ds v_i(s_1, s_2, \dots s_{i-1}, s_{i}', s_{i+1}, \dots s_n) < v_j(s_1, s_2, \dots s_{i-1}, s_{i}', s_{i+1}, \dots s_n).$$

Therefore, as a result of strategies $\ds \sigma_1(s_1), \dots, \sigma_{i-1}(s_{i-1}), \sigma_i(s_i'), \sigma_{i+1}(s_{i+1}), \dots, \sigma_n(s_n)$, buyer $\ds i$ is never offered the good.

Since playing $\ds \sigma_1, \sigma_2, \dots\sigma_n$ is a Nash Equilibrium, buyer $\ds i$ has no incentive to play strategy $\ds \sigma_i(s_i')$ if his or her signal is actually $\ds s_i$, and vice-versa, buyer $\ds i$ has no incentive to play strategy $\ds \sigma_{i}(s_i)$ if his or her signal is actually $\ds s_i'$. This can be rewritten formally as $$\ds p_i\cdot v_i(s_1, \dots s_{i-1}, s_i, s_{i+1}, \dots, s_n) - P_i( \sigma_1(s_1), \dots, \sigma_{i-1}(s_{i-1}), \sigma_i(s_i), \sigma_{i+1}(s_{i+1}), \dots, \sigma_n(s_n) ) \geq $$ $$\ds\geq - P_i( \sigma_1(s_1), \dots, \sigma_{i-1}(s_{i-1}), \sigma_i(s_i'), \sigma_{i+1}(s_{i+1}), \dots, \sigma_n(s_n)) $$ and  $$\ds -  P_i( \sigma_1(s_1), \dots, \sigma_{i-1}(s_{i-1}), \sigma_i(s_i'), \sigma_{i+1}(s_{i+1}), \dots, \sigma_n(s_n)) \geq $$ $$\ds \geq  p_i\cdot v_i(s_1,\dots s_{i-1}, s_{i}', s_{i+1}, \dots s_n) - P_i( \sigma_1(s_1), \dots, \sigma_{i-1}(s_{i-1}), \sigma_i(s_i), \sigma_{i+1}(s_{i+1}), \dots, \sigma_n(s_n) ) .$$ By adding the last two relations we obtain $$\ds p_i\cdot  v_i(s_1,\dots s_{i-1}, s_{i}, s_{i+1}, \dots s_n) \geq p_i \cdot v_i(s_1,\dots s_{i-1}, s_{i}', s_{i+1}, \dots s_n) ,$$ so dividing by $\ds p_i$ and keeping in mind that $\ds v_i$ is increasing in $\ds s_i$, we obtain that $\ds s_i\geq s_{i}'$, which is a contradiction, since we previously chose $\ds s_i' > s_i$. Therefore the lemma is proved $\ds\square$.

Next let us show that our assumption that a buyer $\ds i$'s signal has the same marginal effect on buyer $\ds j$'s valuation for any good is also necessary. We prove this for separable, linear valuation functions, again for simplicity:

\begin{lemma}
Assume there are two buyers and two goods $\ds \{A, B\}$ to be offered, such that each buyer gains no marginal benefit from having both goods, with buyer signal spaces represented as open intervals of the real numbers, $\ds s_{iA} \in\ds\mathcal S_{i, A}, s_{iB}\in \mathcal S_{i, B} $ and positive valuation functions satisfying the following relations for linear functions $\ds w_1, w_2, f_1, f_2, g_1, g_2$ : $$\ds v_{1A} (s_{1A}, s_{2A} ) = w_1(s_{1A}) + g_1(s_{2A}), \mbox{  } v_{1B} (s_{1B}, s_{2B}) = w_1(s_{1B}) + g_2(s_{2B}) ,$$  $$\ds v_{2A} (s_{1A}, s_{2A} ) = w_2(s_{2A}) + f_1(s_{1A}), \mbox{  } v_{2B} (s_{1B}, s_{2B}) = w_2(s_{1B}) + f_2(s_{1B}) ,$$ 

If there is an efficient auction mechanism then $\ds f_1'(\cdot) = f_2'(\cdot)$.
\end{lemma}

Let $\ds (\sigma_1, \sigma_2) $ be a Nash Equilibrium of this auction such that the goods are allocated according to the welfare-maximizing allocation as a result of strategies $\ds (\sigma_1, \sigma_2). $  Fix buyer $\ds 2$'s signals at $\ds s_{2A}$ and $\ds s_{2B}$, and consider the following two sets $\ds T_A$ and $\ds T_B$, defined as
$$\ds T_A = \{ (s_{1A}, s_{1B}) | v_{1A} (s_{1A}, s_{2A}) + v_{2B} (s_{1B}, s_{2B}) \geq  v_{2A} (s_{1A}, s_{2A}) + v_{1B} (s_{1B}, s_{2B}) \} ,$$
$$\ds T_B = \{ (s_{1A}, s_{1B}) | v_{1A} (s_{1A}, s_{2A}) + v_{2B} (s_{1B}, s_{2B}) \leq  v_{2A} (s_{1A}, s_{2A}) + v_{1B} (s_{1B}, s_{2B}) \} .$$

Since this auction is efficient, good $\ds A$ is offered to buyer $\ds 1$ with positive probability if and only if $\ds (s_{1A}, s_{1B}) \in T_A$ and good $\ds B$ is offered to buyer $\ds 1$ with positive probability if $\ds (s_{1A}, s_{1B}) \in T_B$. Assume there exist signal values $\ds (s_{2A}, s_{2B})$ for buyer $\ds 2$ such that the sets $\ds T_A$ and $\ds T_B$ are both nonempty. Otherwise, there is only one possible allocation regardless of the buyers' signals.

Denoting $\ds P_1(t_1) $ buyer $\ds 1$'s expected payment in this auction if he or she plays strategy $\ds \sigma_1$ for signal values $\ds t_1 = (s_{1A}, s_{1B})$ and buyer $\ds 2$ plays $\ds \sigma_2 (s_{2A}, s_{2B})$, first note that $\ds P_1(t_1) = P_1(t_1')$ whenever $\ds t_1$ and $\ds t_1'$ are either both in $\ds T_A$ or both in $\ds T_B$, because the auction is efficient and $\ds (\sigma_1, \sigma_2)$ is a Nash Equilibrium. Thus we can denote $\ds P_A(s_{2A}, s_{2B})$ as buyer $\ds 1$'s payment if he or she bids such that buyer $\ds 1$ receives good $\ds A$, and $\ds P_B(s_{2A}, s_{2B})$ buyer $\ds 1$'s payment if he or she bids such that buyer $\ds 1$ receives good $\ds B$.

Note that for any two signal pairs $\ds t_1 = (s_{1A}, s_{1B}), t_1' = (s_{1A}', s_{1B}') $ such that $\ds t_1 \in T_A$ and $\ds t_1'\in T_B$, we have $$\ds v_{1A}(s_{1A}, s_{2A}) - P_A(s_{2A}, s_{2B}) \geq  v_{1B} (s_{1B}, s_{2B}) - P_B(s_{2A}, s_{2B})$$ and $$\ds v_{1B}(s_{1B}', s_{2B}) - P_B(s_{2A}, s_{2B}) \geq  v_{1A} (s_{1A}', s_{2A}) - P_A(s_{2A}, s_{2B}).$$

This means that $\ds v_{1A} (s_{1A}, s_{2A}) - v_{1B} (s_{1B}, s_{2B}) \geq P_A(s_{2A}, s_{2B}) - P_B(s_{2A}, s_{2B})$ if and only if $\ds (s_{1A}, s_{2A}) \in T_A$, or equivalently $$\ds  w_{1} (s_{1A}) - w_1(s_{1B}) \geq P_A(s_{2A}, s_{2B}) - P_B(s_{2A}, s_{2B}) + g_2(s_{2B}) - g_1(s_{2A})\iff $$ $$\ds \iff w_1(s_{1A}) - w_1(s_{1B}) - (f_1(s_{1A}) - f_2(s_{1B})) \geq w_2(s_{2A}) - w_2(s_{2B}) - (g_1(s_{2A}) - g_2(s_{2B})).$$

Note that the right hand side of both inequalities does not depend on buyer $\ds 1$'s signals. Since we assumed that sets $\ds T_A$ and $\ds T_B$ are both nonempty, and since valuations are linear, there exist infinitely many values $\ds s_{1A}, s_{1B}$ such that we have equality in the relations above. For those values we have that $\ds s_{1A}$ is a linear function of $\ds s_{1B}$ and $\ds f_{1}(s_{1A} ) - f_2(s_{1B})$ is constant, so indeed $\ds f_1'(\cdot) = f_2'(\cdot)$, as desired $\ds\square$.

The two lemmas above prove that our assumptions $\ds (4.2) - (4.3)$ are necessary if we assume separable, linear valuation functions, and thus the only point where we lose generality is when we assume that valuations are separable and linear.

Finally note the following simple and useful lemma:

\begin{lemma}

If propositions $\ds (4.1)$ and $\ds (4.5)$ hold, then the welfare-maximizing allocation is one where each buyer receives at most one good.

\end{lemma}
\emph{Proof.}

Denote $\ds v_{i, S}$ buyer $\ds i$'s valuation for a set $\ds S$ of goods. 

Assume that $\ds M= S_1\sqcup S_2\sqcup \dots S_n$ is a welfare-maximizing allocation, where at least one set has size greater than 1. 

Let $\ds S_i $ be one of the sets of size greater than $\ds 1$, with $\ds j = \Max\limits_{k\in S_i} v_{i, k}$.  Since $\ds \sum _{j=1}^n |S_j| = n$, at least $\ds |S_i| -1$ sets among the $\ds S_1, S_2, \dots S_n$ are empty.

Consider the allocation where buyer $\ds i$ is only assigned good $\ds j$, and the other elements of $\ds S_i$ are distributed to $\ds |S_i|-1$ of the buyers who received nothing in the initial allocation. The overall welfare from this new allocation is at least as big as the previous one, so by continuing this process we will reach a welfare-maximizing allocation where each buyer is assigned at most one good $\square$.

Let us proceed to the efficient auction mechanism. We split the problem in two subsections. Throughout the remaining part of this section we assume that valuation functions satisfy conditions $\ds (4.1) - (4.5) $.

\subsection{There is an equal number of buyers and goods}

\bigskip

Assume there are $\ds n$ buyers and $\ds n$ goods, and that all valuation functions are common knowledge.

Consider the following auction setting (\emph{Auction 1}) :

\begin{enumerate}

\item \emph{Bidding.}  Each buyer $\ds i$ submits an $\ds n$-dimensional vector $\ds s_i = (s_{i1}, s_{i2}, \dots s_{in}) \in \mathbb R^n.$ For every $\ds K\in M$, denote 
\begin{equation}
\ds s_K = (s_{1K}, s_{2K}, \dots , s_{nK})
\end{equation}

\item \emph{Allocation.}  Any permutation $\ds \sigma\in P_n$ defines an allocation $\ds S_{\sigma} = (S_1, S_2, \dots , S_n)$  of the $\ds n$ goods to the $\ds n$ buyers, such that buyer $\ds i$ receives good $\ds \sigma(i)$, or equivalently $\ds S_i = \{\sigma(i)\}$  for all $\ds i\in N$. Under the hypothesis that the $n$-dimensional vectors submitted represent buyers' signals, let $\ds \sigma^{*}$ be the permutation which describes the welfare-maximizing allocation and assign good $\ds \sigma^*(i)$ to buyer $\ds i$, for all $\ds i\in N$. If there are more such possible permutations, uniformly select one of them at random and assign goods according to that permutation.

\item \emph{Payment.} For a given permutation $\ds \sigma\in P_n$, let $\ds S_{\sigma} = (S_1, S_2, \dots S_n)$ be the allocation such that $\ds S_i =\{\sigma(i)\}$ for all $\ds i\in  N$, and let $\ds W(S_{\sigma})$ be the welfare from allocation $\ds S_{\sigma}$, as defined by $\ds (1.4)$. For any $\ds \sigma\in P_n$, define 

\begin{equation}
\ds P_i(\sigma) = \frac{ w_i'(0)}{w_i'(0) - f_i'(0)}\cdot W(S_{\sigma}) - v_{i, \sigma(i)}(s_{\sigma(i)}) -  \frac{ w_i'(0)}{w_i'(0) - f_i'(0)}\cdot\sum_{K\in M} f_i(s_{iK})
\end{equation}

 If the welfare-maximizing allocation from step $\ds (2)$ is described by the permutation $\sigma^*$, then buyer $\ds i$ makes a payment of
\begin{equation}
\ds \Max\limits_{\sigma\in P_n} P_i(\sigma) - P_i(\sigma^*)
\end{equation}
\end{enumerate}
\bigskip
Assume that valuation functions satisfy conditions $\ds (4.1) - (4.5)$, with the assumption that functions $\ds w_i$ and $\ds f_i$ need not be linear, but only satisfy the condition 
\begin{equation}
\ds \mbox{ For any } i\in N, \mbox{ there exist constants }  c_i, d_i \mbox{ such that } w_i (x) = c_i\cdot f_i(x) + d_i, \forall x\in \mathbb R
\end{equation}

We will show that under conditions $\ds (4.1) - (4.5) $, and the linearity of $\ds w_i, f_i$ replaced with condition $\ds (4.9)$, truthful bidding is a Nash Equilibrium and hence the auction is efficient. First, let us prove that the auction is well-defined.

From \emph{Lemma 4.3}, we know that there always exists a welfare-maximizing allocation where each buyer receives exactly one good, therefore the allocation mechanism is well-defined and it does provide the maximal overall welfare if buyers bid truthfully. Let us now deal with the payment function. We prove two lemmas:

\begin{lemma}
If a strictly increasing function $\ds w_i$ satisfies conditions $\ds (4.4)$ and $\ds (4.10)$, then the function $\ds  \frac{ w_i'(x)}{w_i'(x) - f_i(x)}$ is a positive constant.
\end{lemma}

\emph{Proof.} Let $\ds w_i(x) = c_i\cdot f_i(x) + d_i$. This implies that $\ds \frac{ w_i'(x)}{w_i'(x) - f_i(x)} = \frac{c_i}{c_i-1}.$

Since $\ds w_i$ is strictly increasing, we have $\ds c_i\cdot f_i'(x) > 0$ for all $\ds x\in\mathbb R$.

Since $\ds w_i - f_i$ is strictly increasing, we have $\ds (c_i -1)f_1'(x) >0$ for all $\ds x\in\mathbb R$.

By dividing these two relations, we obtain that $\ds \frac{c_i}{c_i -1} >0$, so indeed  $\ds  \frac{ w_i'(x)}{w_i'(x) - f_i(x)}$ is a positive constant $\ds\square$.

\begin{lemma} 
 For any $\ds i\in N $ and for any $\ds\sigma\in P_n$, the value of $\ds P_{i} (\sigma) $ defined in $\ds (4.7)$ does not depend on buyer $\ds i$'s bid $\ds s_i = (s_{i1}, s_{i2}, \dots s_{in})$.
\end{lemma}

Let $\ds \sigma(i) = A$. Then we can rewrite $$\ds P_i(\sigma) = \frac{c_i}{c_i-1}\cdot W(S_{\sigma}) - v_{i, A} (s_A) -\frac{c_i}{c_i-1} \cdot\sum_{K\in M} f_i(s_{iK}).$$

Note that for $\ds K\neq A$, the coefficient of $\ds f_i(s_{iK}) $ is zero in the expression above, since the first term of the right hand side provides a coefficient of $\ds \frac{c_i}{c_i -1}$, while the third term provides a coefficient of $\ds -\frac{c_i}{c_i-1}$ . Let's look at all terms involving $\ds s_{iA}$: The first term of the right hand side gives $\ds \frac{c_i}{c_i -1} w_i(s_{iA}) $, the second term $\ds - w_i(s_{iA})$ and the third term $\ds - \frac{c_i}{c_i -1} \cdot f_i(s_{iA})$. Therefore, the sum of all the functions involving $\ds s_{iA}$ in the expression $\ds P_i(\sigma)$ is $$\ds \frac{c_i}{c_i-1}w_i(s_{iA}) - w_i(s_{iA}) - \frac{c_i}{c_i-1} f_i(s_{iA})= \frac{w_i(s_{iA}) - c_i f_i(s_{iA})}{c_i-1}=\frac{d_i}{c_i-1},$$ which is indeed a constant.

Therefore the value of $\ds P_i (\sigma)$ does not depend on buyer $\ds i$'s bid $\ds s_i = (s_{i1}, s_{i2}, \dots s_{in})$ for any $\ds i\in N, \sigma\in P_n$ $\ds\square$.

We are now ready to prove that this auction is efficient.

\begin{theorem}
If valuation functions satisfy conditions $\ds (4.1) - (4.5)$, and the linearity of each $\ds w_i$ and $\ds f_i$ is replaced by $\ds (4.9)$, then truthful bidding, i.e. bidding one's own signal values, represents a Nash Equilibrium of this auction.
\end{theorem}

Fix a buyer $\ds i$ and assume that every buyer $\ds j\neq i$ bids truthfully, i.e. bids his or her actual signals for the $n$ goods. We prove that buyer $\ds i$ has no incentive to deviate from truthful bidding.

Let $\ds \sigma_1$ be the welfare-maximizing permutation of the goods to buyers given their true signals, and let $\ds \sigma_2 = \argmax\limits_{\sigma\in P_n} P_i(\sigma)$. If buyer $\ds i$ bids truthfully, the allocation described by permutation $\ds \sigma_1$ is selected, and buyer 1 makes a payment of $\ds P_i(\sigma_2) - P_i(\sigma_1)$, thus his or her overall utility is $\ds v_{i, \sigma_1(i)}(s_{\sigma_1(i)}) + P_i (\sigma_1) - P_i(\sigma_2).$ Using definition $\ds (4.7)$, this can be rewritten as 
\begin{equation}
\ds \frac{c_i}{c_i-1}(W(S_{\sigma_1}) - W(S_{\sigma_2})) + v_{i, \sigma_2(i)}(s_{\sigma_2(i)})
\end{equation}

Since $\ds \sigma_1$ is the welfare-maximizing allocation, and since valuations are nonnegative, the utility obtained from truthful bidding is  nonnegative for buyer $\ds i$, so he or she has no incentive to not participate in the auction.

Now assume that buyer bids so that goods are allocated according to a permutation $\ds \sigma_3$. In this case, he or she makes a payment of $\ds P_i(\sigma_2) - P_i(\sigma_3)$ and obtains a utility of $$\ds \frac{c_i}{c_i-1} ( W(S_{\sigma_3}) - W(S_{\sigma_2}) ) + v_{i, \sigma_2(i)}(s_{\sigma_2(i)}).$$

Therefore, the difference in utility between receiving allocation $\ds\sigma_1$ and $\ds\sigma_3$ is $\ds \frac{c_i}{c_{i}-1}( W(S_{\sigma_1}) - W(S_{\sigma_2}) ) \geq 0$, since $\ds \sigma_1$ is the welfare-maximizing allocation. 

Therefore buyer $\ds i$ has no incentive to deviate to bidding such that another allocation is selected, and since the payment does not depend on his bid, we obtain that truthful bidding is a best response, and therefore truthful bidding consists of a Nash Equilibrium. Since the allocation mechanism assigns according to the welfare-maximizing allocation, this auction mechanism is indeed efficient $\ds\square$ .

\bigskip

Let's apply this mechanism to Example (2) . Recall that there are two goods to be auctioned, and that valuations are described by 

\begin{equation}
\ds v_{1A}(s_{1A}, s_{2A}) = s_{1A} +\frac{1}{2}s_{2A},\mbox{   } v_{1B}(s_{1B}, s_{2B}) = s_{1B} +\frac{1}{2}s_{2B}
\end{equation}
\begin{equation}
\ds v_{2A}(s_{1A}, s_{2A}) = s_{2A} +\frac{1}{3}s_{1A},\mbox{   }  v_{2B}(s_{1B}, s_{2B}) = s_{2B} +\frac{1}{3}s_{1B}
\end{equation}

Say that buyer $\ds 2 $ has signals $\ds s_{2A} = 2, s_{2B} =4$, and let $\ds s_{1A}\geq -1, s_{1B}\geq -2$ be buyer 1's signals.  If buyer $\ds 2 $ bids truthfully, and buyer $\ds 1$ bids $\ds (s_{1A} ', s_{1B} ')$, then from the auction designer's perspective, the apparent welfare from allocation $\ds (A, B)$, where buyer 1 gets A and buyer 2 gets B, is $\ds s_{1A}' + \frac{1}{3} s_{1B}' + 5$ and the apparent welfare from allocation $\ds (B, A)$, where buyer $\ds 1$ gets B and buyer $\ds 2$ gets A is $\ds \frac{1}{3}s_{1A}' + s_{1B} +4$. 

Allocation $\ds (A, B)$ is selected if $\ds s_{1B} ' - s_{1A} ' < \frac{3}{2}$, allocation $\ds (B, A)$ is selected if $\ds s_{1B} ' - s_{1A} ' > \frac{3}{2},$ and in case  $\ds s_{1A} - s_{1B} = \frac{3}{2}$, each allocation has a $\ds\frac{1}{2}$ chance of being selected. 

Let $\ds \sigma_1 $ represent allocation $\ds (A, B)$ and $\ds \sigma_2 $ represent allocation $\ds (B, A)$. Then $\ds P_1(\sigma_1) = \frac{13}{2}, P_1(\sigma_2) = 4$, and therefore buyer $\ds 1$ faces the following decision :

\begin{enumerate}

\item If $\ds s_{1B}' - s_{1A}' <\frac{3}{2}$, buyer 1 makes a payment of zero, and obtains utility $\ds s_{1A} +1 $ .

\item If $\ds s_{1B}' - s_{1A}' >\frac{3}{2}$, buyer 1  makes a payment of $\ds \frac{13}{2}-4 =\frac{5}{2}$, and obtains utility $\ds s_{1B} -\frac{1}{2}$.

\item If $\ds s_{1B}' - s_{1A}' =\frac{3}{2}$, buyer 1 makes an expected payment of $\ds\frac{5}{4}$, and obtains expected utility $\ds \frac{s_{1A} + s_{1B}}{2} -\frac{1}{4}$.

\end{enumerate}

Note from above that buyer $\ds 1$ is strictly better off from bidding $\ds (1)$ if and only if $\ds s_{1B} - s_{1A} <\frac{3}{2}$, that he or she is strictly better off from bidding $\ds (2)$ if and only if $\ds s_{1B} - s_{1A} >\frac{3}{2}$, and that he or she is indifferent to the three strategies if and only if $\ds s_{1B} - s_{1A} =\frac{1}{3}$. 

Since bidding $\ds s_{1A}' = s_{1A}$ and $\ds s_{1B}' = s_{1B}$ satisfies the conditions above, it is indeed a best response: there is no other bidding strategy that could make buyer 1 better off. 

Let us now move on to the second case.

\subsection {There are strictly more buyers than goods}

\bigskip

Assume there are $\ds n$ buyers and $\ds m<n$ goods, and that all valuation functions are common knowledge.

Consider the following auction setting:

\begin{enumerate}

\item \emph{Bidding.} Each buyer $\ds i$ submits an $\ds m$- dimensional vector $\ds s_i = (s_{i1}, s_{i2}, \dots s_{im}) \in \mathbb R^m$ .For every $\ds K\in M$, denote 
\begin{equation}
\ds s_K = (s_{1K}, s_{2K}, \dots , s_{nK})
\end{equation}

\item \emph{Allocation.} Given the vectors that the buyers submitted, the goods are allocated according to the welfare-maximizing allocation $\ds (S_1, S_2, \dots S_n)$ under the hypothesis that for every buyer, the vector that he or she submitted represents his or her signal values, and under the condition that each buyer receives at most one good.  If such an allocation is not unique, uniformly select one of the welfare-maximizing allocations at random.

\item \emph{Payment.} Assume that the above allocation assigns each good $\ds K \in M $ to bidder $\ds i_K$, such that for every $\ds A,B \in M$ with $ A\neq B,$ we have $\ds i_A\in N, i_B\in N, i_A\neq i_B$. For a given buyer $\ds i$, the following mechanism describes his or her payment:

i) If buyer $\ds i$ is not assigned any good, buyer $\ds i$ makes no payment

ii) If buyer $\ds i$ is assigned good $\ds K$, consider the welfare optimizing allocation of goods from the set $\ds M$ to the buyers in the set $\ds N - \{i\}$ according to the same rules as in step $\ds (2)$ above. Assume that under this new allocation, good $\ds A$ is assigned to buyer $\ds j_A\neq i$, such that for every $\ds A,B \in M$ with $ A\neq B,$ we have $\ds j_A\in N, j_B\in N, j_A\neq j_B$. Buyer $\ds i$ pays

\begin{equation}
\ds v_{i, K}(s_{1K}, s_{2K}, \dots ,s_{(i-1) K}, s_{iK}^*, s_{(i+1) K}, \dots ,s_{nK}),
\end{equation}
 where $\ds s_{iK}^*$ solves the following equation:
$$\ds
 v_{i, K}(s_{1K}, s_{2K}, \dots ,s_{(i-1) K}, s_{iK}^*, s_{(i+1) K}, \dots ,s_{nK}) + \sum_{A\in M, A\neq K} v_{i_A, A} (s_A) = $$

\begin{equation}
v_{j_K, K}(s_{1K}, s_{2K}, \dots ,s_{(i-1) K}, s_{iK}^*, s_{(i+1) K}, \dots s_{nK}) + \sum_{A\in M, A\neq K} v_{j_A, A}(s_A),
\end{equation}
 where $\ds  s_A = (s_{1A}, s_{2A}, \dots , s_{nA})$, for any $\ds A\in M$, as in relation $\ds (4.13)$

\end{enumerate}

We claim that truthful bidding is a Nash Equilibrium of this auction. In turn, this would imply that the auction is efficient, given the allocation mechanism above.

First, let us prove that the auction is well-defined.

From lemma $\ds (4.3)$, we know that we will always find an allocation satisfying condition $(2)$ of our auction. Let us next prove that the payment mechanism is well-defined.

\begin{lemma}

Equation $\ds  (4.14) $ has a unique solution $\ds s_{iK}^*$ and the payment

$\ds v_{i, K}(s_{1K}, s_{2K}, \dots ,s_{(i-1) K}, s_{iK}^*, s_{(i+1) k}, \dots s_{nK})$ does not depend on buyer $\ds i$'s bid $\ds s_i = (s_{i1}, s_{i2}, \dots s_{im})$ .
\end{lemma}

\emph{Proof.} Using the linearity condition $\ds (4.2)$, relation $\ds (4.14)$ can be rewritten as $$\ds w_i(s_{iK}^*) +\sum _{t\neq i} f_t(s_{tK}) + \sum_{A\in M, A\neq K} v_{i_A, A}(s_A) = $$ $$\ds  w_{j_K} (s_{j_KK}) + f_i(s_{iK}^*) + \sum_{i\neq t\neq j_K} f_{t}(s_{tK}) + \sum_{A\in M, A\neq K} v_{jA, A}(s_A) ,$$ or equivalently 

\begin{equation}
\ds w_i(s_{iK}^*) + f_{j_K}(s_{j_KK}) - w_{j_K}(S_{j_KK}) - f_i(s_{iK}^*) = \sum_{A\in M, A\neq K} \left(v_{i_A, A}(s_A) - v_{j_A, A}(s_A)\right)
\end{equation}

Note again from the linearity condition $\ds (4.2)$ that even though $\ds v_{iA, A} (s_A)$ depends on $\ds s_{iA}$ because of the additive term $\ds f_i(s_{iA})$, the difference $\ds v_{i_A, A}(s_A) - v_{j_A, A}(s_A)$ does not. Therefore, neither the left nor the right hand side of relation $\ds (3.8)$ depend on any of the signals $\ds s_{i1}, s_{i2}, \dots s_{im}$.

Moreover, relation $\ds (4.16)$ is linear in $\ds \ds s_{iK}^*$, so it indeed has a unique solution $\ds\square$.

Let us prove another useful lemma:

\begin{lemma}

The welfare optimizing allocation of the goods in $\ds M$ to the buyers $\ds N - \{i\}$ does not depend on buyer $\ds i$'s signal values, even though individual valuations do.

\end{lemma}

\emph{Proof.} Let  $\ds  s_A = (s_{1A}, s_{2A}, \dots , s_{nA})$, for any $\ds A\in M$, as in relation $\ds (4.13)$. Consider two allocations, one where every good $\ds K\in M$ is assigned to buyer $\ds i_K\in N-\{i\}$, such that for all $\ds A, B\in M, A\neq B$ we have $\ds  i_A\neq i_B$, and another allocation  where every good $\ds K\in M$ is assigned to buyer $\ds j_K\in N-\{i\}$, such that for all $\ds A, B\in M, A\neq B$ we have $\ds  j_A\neq j_B$.

The difference in welfare between the two allocations is $$\ds \sum_{K\in M} v_{i_K, K}(s_K) - \sum_{K\in M} v_{j_K, K}(s_K) =\sum_{K\in M} \left( v_{i_K, K}(s_K) - v_{j_K, K}(s_K)\right).$$ Since $\ds j_K\neq i\neq i_K$ for all $\ds K\in M$, we obtain from condition $(4.2)$ that each term $\ds v_{i_K, K}(s_K) - v_{j_K, K}(s_K)$ of the sum above does not depend on $\ds s_{iK}$. Therefore the welfare ordering between allocations of the goods in $\ds M$ to buyers in $\ds N -\{i\}$ does not depend on buyer $\ds i$'s signal, as desired $\ds\square$.

We are now ready to prove our main result.

\begin{theorem}
Truthful bidding, i.e. bidding one's own signal values for every good is a Nash Equilibrium under the auction design presented above.
\end{theorem}

\emph{Proof.} Let $\ds i \in N$ be one of the buyers, and assume that every other buyer $\ds j\neq i$ bids truthfully, i.e. for all $\ds K\in M$,  the signal $\ds s_{jK}$ that buyer $\ds j$ submits is actually buyer $\ds j$'s signal value for good $\ds K$.

Let us prove that truthful bidding is a best response for buyer $\ds i$ .

We encounter two cases:

\begin{enumerate}

\item If buyer $\ds i$ bids truthfully, he or she obtains a good.

\item  If buyer $\ds i$ bids truthfully, he or she does not obtain any of the goods.

\end{enumerate}
Let us deal with Case 1 first:

First assume that the welfare-maximizing allocation of goods $\ds M$ to buyers $\ds M -\{i\}$ is such that buyer $\ds j_A$ receives good $\ds A$, for all goods $\ds A\in M$.

Assume that under truthful bidding, buyer $\ds i$ receives good $\ds K$, while for all $\ds A\neq K$, buyer $\ds i_A$ receives good $\ds A$. Since all buyers bid truthfully, this represents the ex-ante welfare-maximizing allocation.

In this case, buyer $\ds i$ receives good $\ds K$ and has to pay 

$\ds v_{iK}(s_{1K}, s_{2K}, \dots ,s_{(i-1) K}, s_{iK}^*, s_{(i+1) K}, \dots s_{nK})$, where $\ds s_{iK}^*$ solves $$ \ds w_i(s_{iK}^*) + f_{j_K}(s_{j_KK}) - w_{j_K}(S_{j_KK}) - f_i(s_{iK}^*) = \sum_{A\in M, A\neq K} \left(v_{i_A, A}(s_A) - v_{j_A, A}(s_A)\right).$$

Therefore buyer $\ds i$'s utility in this auction is his or her valuation minus the payment, which simplifies to $\ds w_i(s_{iK}) - w_i(s_{iK}^*)$ because of the linearity condition $\ds (4.2)$.

We will prove that no bidding for buyer $\ds i$ can result in a higher utility. We will consider two subcases:

 i. buyer $\ds i$ bids so that he or she does not receive any goods

ii. buyer $\ds i$ bids so that he or she receives good $\ds L$ (which may or may not be the same as good $\ds K$, the one obtained from truthful bidding)

We deal with these cases one by one.

\bigskip
Consider $$\ds f(t) = v_{i, K}(s_{1K}, s_{2K}, \dots ,s_{(i-1) K}, s_{iK} - t, s_{(i+1) K}, \dots s_{nK}) + \sum_{A\in M, A\neq K} v_{i_A, A} (s_A) -$$ $$ -\ds  v_{j_K, K}(s_{1K}, s_{2K}, \dots ,s_{(i-1) K}, s_{iK}-t, s_{(i+1) K}, \dots s_{nK}) -\sum_{A\in M, A\neq K} v_{j_A, A}(s_A)$$
\bigskip
Observe that we can use relation $\ds (4.16)$ to simplify $\ds f(t)$ to  $$ \ds f(t) = \ds w_i(s_{iK} - t) + f_{j_K}(s_{j_KK}) - w_{j_K}(S_{j_KK}) - f_i(s_{iK}-t)  + \sum_{A\in M, A\neq K} \left(v_{i_A, A}(s_A) - v_{j_A, A}(s_A)\right).$$

Note that $\ds f'(t) = f_i'(s_{iK}-t) -  w_i'(s_{iK}-t) <0 $ for all $\ds t\in\mathbb R$  so from relation $\ds (4.4)$, $\ds f$ is decreasing.

Also observe that $\ds f(s_{iK}- s_{iK}^*) =0 $.

Finally, note that $\ds f(0)= W_i - W_j$, where $\ds W_i$ is the welfare obtained when good $\ds K$ is assigned to buyer $\ds i$ and every good $\ds A\neq K$ is assigned to buyer $\ds i_A$, and $\ds W_j$ is the welfare obtained by allocating the goods in $\ds M$ to buyers $\ds N-\{i\}$ such that good $\ds A\in M$ is assigned to buyer $\ds  j_A$.

Since $\ds W_i$ is the welfare obtained from truthful bidding, it is higher than all other welfares, so $\ds f(0) \geq 0$, and therefore $\ds f(0) \geq f(s_{iK}- s_{iK}^*) $. But $\ds f$ is decreasing, so $\ds s_{iK} \geq s_{iK}^*$, which implies $\ds w_i(s_{iK}) - w_i(s_{iK}^*) \geq 0$.

Let us deal with the 2 subcases now:

\emph{Subcase i}. buyer $\ds i$ bids so that he or she does not receive any goods.

In this case buyer $\ds i$'s utility is zero, which is not greater than the utility obtained by truthful bidding, $\ds w_i(s_{iK}) - w_i(s_{iK}^*)$. Therefore buyer $\ds i$ does not have an incentive to deviate to this case $\ds\square$.

\emph{Subcase ii.} Assume buyer $\ds i$ sends signals $\ds s_i' =(s_{i1}', s_{i2}', \dots , s_{im}')$, such that the resulting allocation gives good $\ds L$ to buyer $\ds i$ and good $\ds A\in M$ to buyer $\ds i'_A$, for all $\ds A\neq L$. 

Therefore, buyer $\ds i$ receives good $\ds L$ and has to pay 

$\ds v_{iL}(s_{1L}, s_{2L}, \dots ,s_{(i-1) L}, s_{iL}^{**}, s_{(i+1) L}, \dots s_{nL})$, where $\ds s_{iL}^{**}$ solves $$ \ds w_i(s_{iL}^{**}) + f_{j_{L}}(s_{j_{L}L}) - w_{j_{L}}(S_{j_{L}L}) - f_i(s_{iL}^{**}) = \sum_{A\in M, A\neq L} \left(v_{i_A', A}(s_A) - v_{j_A, A}(s_A)\right).$$

Therefore, if buyer $\ds i$ bids truthfully, he or she obtains utility $\ds w_i(s_{iK}) - w_i(s_{iK}^*)$, while in this case he or she obtains utility $\ds w_i(s_{iL}) - w_i(s_{iL}^{**})$ . 

Consider $$\ds g(t) = v_{i, L}(s_{1L}, s_{2L}, \dots ,s_{(i-1) L}, s_{iL} - t, s_{(i+1) L}, \dots s_{nL}) + \sum_{A\in M, A\neq L} v_{i_A', A} (s_A) -$$ $$ -\ds  v_{j_{L}, L}(s_{1L}, s_{2L}, \dots ,s_{(i-1) L}, s_{iL}-t, s_{(i+1) L}, \dots s_{nL}) -\sum_{A\in M, A\neq L} v_{j_A, A}(s_A)$$
\bigskip
Observe that we can use relation $\ds (4.16)$ to simplify $\ds g(t)$ to  $$\ds  g(t) = \ds w_i(s_{iL} - t) + f_{j_L}(s_{j_LL}) - w_{j_L}(S_{j_LL}) - f_i(s_{iL}-t)  + \sum_{A\in M, A\neq L} \left(v_{i_A', A}(s_A) - v_{j_A, A}(s_A)\right).$$

Note that $\ds g'(t) = f_i'(s_{iL}-t) -  w_i'(s_{iL}-t) <0 $ for all $\ds t\in\mathbb R$  so from relation $\ds (4.4)$, $\ds g$ is decreasing. Moreover, since the functions $\ds f_i$ and $\ds w_i$ are linear, we obtain $\ds g'(t) = f'(t)$ for every $\ds t\in \mathbb R$, so the function $\ds f(t) - g(t)$ is constant.

Also observe that $\ds g(s_{iL}- s_{iL}^{**}) =0 $.

Finally, note that $\ds g(0)= W_{i'} - W_j$, where $\ds W_{i'}$ is the welfare obtained when good $\ds L$ is assigned to buyer $\ds i$ and every good $\ds A\neq L$ is assigned to buyer $\ds i'_A$, and $\ds W_j$ is the welfare obtained by allocating the goods in $\ds M$ to buyers $\ds N-\{i\}$ such that good $\ds A\in M$ is assigned to buyer $\ds  j_A$.

Since truthful bidding results in the welfare-maximizing allocation, we have $\ds W_i \geq W_{i'},$ so $\ds f(0) - g(0) = W_{i}- W_{i'} \geq 0$. Therefore $\ds f(t) \geq g(t)$ for all $\ds t\in \mathbb R$.

This implies that $$\ds f(s_{iL} - s_{iL}^{**}) \geq g( s_{iL} - s_{iL}^{**}) =0 = f(s_{iK}- s_{iK}^*),$$ so $\ds s_{iK} - s_{iK}^* \geq s_{iL} - s_{iL}** ,$ which in turn implies $$\ds w_i(s_{iK}) - w_i(s_{iK}^*) \geq w_{i}(s_{iL})- w_i(s_{iL}^**),$$ therefore buyer $\ds i$ has no incentive to deviate to this case $\ds\square$.

Therefore if buyer $\ds i$ bids truthfully and he or she is assigned a good, there is no strategy which would make him or her better off.

Now we deal with the second case:

\emph{Case 2.} If buyer $\ds i$ bids truthfully and he or she does not obtain any of the goods.

In this case, if buyer $\ds i$ bids truthfully, he or she obtains zero utility. We shall prove that no other strategy gives a higher utility. The reasoning is very similar to \emph{subcase 2} above.

Assume buyer $\ds i$ sends signals $\ds s_i' =(s_{i1}', s_{i2}', \dots , s_{im}')$, such that the resulting allocation gives good $\ds L$ to buyer $\ds i$ and good $\ds A\in M$ to buyer $\ds i'_A$, for all $\ds A\neq L$. 

Therefore, buyer $\ds i$ receives good $\ds L$ and has to pay 

$\ds v_{iL}(s_{1L}, s_{2L}, \dots ,s_{(i-1) L}, s_{iL}^{**}, s_{(i+1) L}, \dots s_{nL})$, where $\ds s_{iL}^{**}$ solves $$ \ds w_i(s_{iL}^{**}) + f_{j_{L}}(s_{j_{L}L}) - w_{j_{L}}(S_{j_{L}L}) - f_i(s_{iL}^{**}) = \sum_{A\in M, A\neq L} \left(v_{i_A', A}(s_A) - v_{j_A, A}(s_A)\right).$$

Therefore, if buyer $\ds i$ bids truthfully, he or she obtains utility $\ds 0$, while in this case he or she obtains utility $\ds w_i(s_{iL}) - w_i(s_{iL}^{**})$ . 

Consider $$\ds h(t) = v_{i, L}(s_{1L}, s_{2L}, \dots ,s_{(i-1) L}, s_{iL} - t, s_{(i+1) L}, \dots s_{nL}) + \sum_{A\in M, A\neq L} v_{i_A', A} (s_A) -$$ $$ -\ds  v_{j_{L}, L}(s_{1L}, s_{2L}, \dots ,s_{(i-1) L}, s_{iL}-t, s_{(i+1) L}, \dots s_{nL}) -\sum_{A\in M, A\neq L} v_{j_A, A}(s_A)$$
\bigskip
Observe that we can use relation $\ds (4.16)$ to simplify $\ds h(t)$ to  $$ h(t) = \ds w_i(s_{iL} - t) + f_{j_L}(s_{j_LL}) - w_{j_L}(S_{j_LL}) - f_i(s_{iL}-t)  + \sum_{A\in M, A\neq L} \left(v_{i_A', A}(s_A) - v_{j_A, A}(s_A)\right).$$

Note that $\ds h'(t) = f_i'(s_{iL}-t) -  w_i'(s_{iL}-t) <0 $ for all $\ds t\in\mathbb R$, from relation $\ds (4.4)$, so $\ds h$ is decreasing.

Also observe that $\ds h(s_{iL}- s_{iL}^{**}) =0 $.

Moreover, $\ds h(0) = W_{i'} - W_j$, where $\ds W_{i'}$ is the welfare obtained if good $\ds L$ is assigned to buyer $\ds i$ and good $\ds A\neq L$ is assigned to buyer $\ds i_A$, and $\ds W_j$ is the welfare obtained by allocating the goods in $\ds M$ to buyers $\ds N- \{i\}$ such that good $\ds A$ is assigned to buyer $\ds j_A$. Since truthful bidding provides the welfare-maximizing allocation, since truthful bidding assigns no goods to buyer $\ds i$, and since $\ds W_j$ is the highest possible welfare obtained from allocating goods in $\ds M$ to buyers $\ds N - \{i\}$, we obtain that $\ds W_j$ is the highest possible welfare obtained from allocating goods in $\ds M$ to all the $\ds N$ buyers. Therefore $\ds W_{i'} - W_j \leq 0$, so $\ds h(0) \leq 0$. This implies that 

$\ds h(0) \leq h(s_{iL}- s_{iL}^{**}) $, so $\ds s_{iL} \leq s_{iL}^{**}$, which finally implies that 

$\ds w_i(s_{iL}) - w_i(s_{iL}^{**}) \leq 0,$ therefore buyer $\ds i$ has no incentive to deviate from truthful bidding, since he or she will not obtain a higher utility $\ds \square$.

Therefore Theorem 4.7 is proved, and the auction mechanism is indeed efficient.

Let us consider an application of this auction, for $3$ buyers and $\ds 2$ goods, $\ds A, B$. Assume that valuation functions can be described as 

\begin{equation}
v_{1A}(s_{1A}, s_{2A}, s_{3A}) = s_{1A} + \frac{1}{2}s_{2A}+\frac{1}{3} s_{3A}, \mbox{  } v_{1B}(s_{1B}, s_{2B}, s_{3B}) = s_{1B} +\frac{1}{2}s_{2B}+\frac{1}{2}s_{3B}
\end{equation}

\begin{equation}
v_{2A}(s_{1A}, s_{2A}, s_{3A}) = \frac{1}{2}s_{1A} + s_{2A}+\frac{1}{3} s_{3A}, \mbox{  } v_{2B}(s_{1B}, s_{2B}, s_{3B}) = \frac{1}{2}s_{1B} +s_{2B}+\frac{1}{2}s_{3B}
\end{equation}

\begin{equation}
v_{3A}(s_{1A}, s_{2A}, s_{3A}) = \frac{1}{2}s_{1A} + \frac{1}{2}s_{2A}+ s_{3A}, \mbox{  } v_{3B}(s_{1B}, s_{2B}, s_{3B}) = \frac{1}{2}s_{1B} +\frac{1}{2}s_{2B}+s_{3B}
\end{equation}

Let $\ds s_{2A} =2, s_{2B}=2, s_{3A}=3, s_{3B}=6 $, and $\ds s_{1A}, s_{1B} >0 $. Note that these valuation functions satisfy our conditions $\ds (4.1) - (4.5)$. Using the specific examples above, the valuations can be rewritten as

$$\ds v_{1A}(s_{1A}) = s_{1A} + 2, v_{1B}(s_{1B}) = s_{1B} + 3$$

$$\ds v_{2A}(s_{1A}) = \frac{1}{2} s_{1A} +3, v_{2B}(s_{1B}) = \frac{1}{2} s_{1B} + 4$$

$$\ds v_{3A}(s_{1A}) =\frac{1}{2} s_{1A} + 4, v_{3B}(s_{1B}) = \frac{1}{2}s_{1B} + 7$$

We describe an allocation by $ (X_1, X_2, X_3)$ if buyer $\ds i$ receives the goods in set $\ds X_i$. Assume that buyers 2 and 3 bid truthfully, and that buyer $\ds 1$ bids $\ds (s_{1A}', s_{1B}')$. Given the valuations above, there are 3 possible allocations which maximize welfare from the auction designer's point of view:

\begin{enumerate}
\item $\ds (A,\emptyset, B)$ if $\ds s_{1A}' \geq \Max(2, s_{1B}' -4)$.

\item $\ds (B,\emptyset, A)$ if $\ds s_{1B}'\geq \Max(6, s_{1A}' -4) $.

\item $\ds (\emptyset, A, B)$ if $\ds s_{1A}' \leq 2, s_{1B}'\leq 6$.
\end{enumerate}

Note that the welfare-maximizing allocation of the 2 goods to buyers $\ds \{2, 3\}$ is $\ds (\emptyset, A, B)$.

If buyer $\ds 1$ bids so that he or she receives good $\ds A$, then the value of $\ds s_{1A}^*$ from relation $\ds (4.15)$ is $\ds 2$, and buyer $\ds 1$ has to make a payment of $\ds v_{1A} (2) =4$.

If buyer $\ds 1 $ bids so that he or she receives good $\ds B$, then the value of $\ds s_{1B}^*$ from relation $\ds (4.15)$ is $\ds 6$, and buyer $\ds 1$ has to make a payment of $\ds v_{1B}(6) =9$.

Therefore, buyer $\ds 1$ has to choose between the following allocations and payments:

\begin{enumerate}

\item If allocation (1) is selected, buyer $\ds 1$ gets good $\ds A$ and obtains a utility of $\ds s_{1A} -2 $.

\item If allocation (2) is selected, buyer $\ds 1$ gets good $\ds B$ and obtains a utility of $\ds s_{1B} -6$.

\item If allocation (3) is selected, buyer $\ds 1$ gets nothing and obtains a utility of zero.

\end{enumerate}

We observe that allocation $\ds (1)$ provides the highest utility for buyer $\ds 1$ if and only if $\ds s_{1A} \geq \Max(2, s_{1B} -4)$, allocation $\ds (2)$ provides the highest utility for buyer $\ds 1$ if and only if  $\ds s_{1B}\geq \Max(6, s_{1A} -4) $, and allocation $\ds (3)$ provides the highest utility for buyer $\ds 1$ if and only if  $\ds s_{1A} \leq 2, s_{1B}\leq 6$. Note that these conditions are symmetric to the allocation rules presented previously, and therefore truthful bidding is indeed optimal, so this auction is efficient.

Note that in this section we have assumed that the auction designer is aware of the functional forms of buyer valuations. We show that this condition is not necessary, by presenting an auction mechanism where the designer does not know buyer functional forms. They are only known amongst each buyer.

\section{Common values and multidimensional signals with no prior knowledge from the auction designer}

Assume, as in the previous section, that there is a set of $\ds M$ goods to be distributed to a group of $\ds N$ buyers, and that assumptions $\ds (4.1) - (4.5)$ and $\ds (4.9)$ hold. The buyers observe their own signals, but do not know the other buyers' signals. However, every buyer knows every other buyer's valuation functions. Since each valuation is linear, if a buyer $\ds i$ knew the specific valuation that every other buyer had, he or she would obtain a system of equations with information about other buyers' valuation components. This could lead the buyer to discover the relevant components of his or her own valuation function. This rationalle, if properly applied, can be used to construct a mechanism where the true valuation functions and signals are revealed under truthful bidding, without requiring the auction designer to know any valuation function of the buyers. In fact, the only thing that the auction designer knows is that buyer valuations satisfy conditions $\ds (4.2)$ and $\ds (4.9)$. These assumptions are not necessary, but they will simplify the mechanism, and since the buyers do in fact have separable valuations, it does not seem farfetched to assume that the auction designer is aware of that.

This section builds upon the previous 2 sections, using a similar method as \cite{maskin00}.  

Assume for simplicity that $\ds \mathcal S_{K} = \mathbb R^{n}$, for all $\ds K\in M$, i.e. every signal can take on any real value, for any good and any buyer. 

Once again we assume that valuation functions are separable and that the marginal effect that every buyer has on the valuations of other buyers is independent of the good or of the buyer, i.e. there exist functions $\ds w_i, f_i$ such that for all $\ds i\in N$ and for all $\ds K\in M$,

\begin{equation}
v_{i, K} (s_{1K}, s_{2K},\dots ,s_{nK} ) = w_i(s_{iK}) +\sum_{j\neq i} f_j(s_{jK}), \forall \ds K\in M, \forall (s_{1K}, s_{2K}, \dots s_{nK})\in \mathcal S_K
\end{equation}

Moreover, assume as before that each function $\ds w_i$ is a linear transformation of $\ds f_i$, i.e. there exist constants $\ds c_i, d_i$ such that $\ds c_i >1$ and

\begin{equation}
w_i(x) = c_if_i(x)+d_i, \mbox{ for all } x\in \mathbb R.
\end{equation}

Fix a good $\ds A\in M$ and a buyer $\ds i \in N$. Assume that each buyer $\ds j$ has a valuation function $\ds v_{j,A}$. We say that a linear function $\ds b_{iA} :\mathbb R^{n-1} \mapsto \mathbb R$ represents truthful bidding for buyer $\ds i$ with signal $\ds s_{iA}$ if there exist real numbers $\ds x_{i1}, x_{i2}, \dots x_{in}$ such that for any $\ds v_{-i} = (v_1, v_2, \dots, v_{i-1}, v_{i+1}, \dots, v_{n})\in \mathbb R^{n-1}$, we have
 
\begin{equation}
\ds b_{iA} (v_{-i}) = x_{ii} + \sum_{j\neq i} x_{ij}\cdot v_j 
\end{equation}
and 

\begin{equation}
b_{iA}(v_{1A}(s_A), v_{2A}(s_A), \dots, v_{(i-1)A}(s_A), v_{(i+1)A}(s_A), \dots, v_{nA}(s_A)) = v_{iA}(s_A),
\end{equation}

for all vectors $\ds s_A= (s_{1A}, s_{2A}, \dots, s_{(i-1)A}, s_{iA}, \dots , s_{nA}) \in\mathbb R^n$ . Note that $\ds s_{iA}$ is fixed, while all other signal values are not.

The following lemma will be very useful:

\begin{lemma}
Fix a good $\ds A\in M$, and a buyer $\ds i\in N$ with signal $\ds s_{iA}$. If buyer valuations satisfy relations $\ds (5.1)$ and $\ds (5.2)$, and if functions $\ds f_j$ are continuous and nonconstant for all $\ds j\neq i, j\in N$, then a truthful bidding function satisfying conditions $\ds (5.3)$ and $\ds (5.4)$ exists and is unique.
\end{lemma}

\emph{Proof.} Assume there exists a truthful bid function satisfying conditions $\ds (5.3)$ and $\ds (5.4)$. Then condition $\ds (5.4)$ can be rewritten as $$\ds x_{ii} + \sum_{j\neq i} x_{ij} \cdot\left( w_{j}(s_{jA}) + \sum _{t\neq j} f_t(s_{tA})\right) = w_{i}(s_{iA}) +\sum_{j\neq i } f_{j}(s_{jA}),$$ or equivalently 
\begin{equation}
\ds x_{ii} + \sum_{j\neq i} x_{ij} \cdot\left( c_j\cdot f_{j}(s_{jA}) + d_j + \sum _{t\neq j} f_t(s_{tA})\right) = c_{i}\cdot f_i(s_{iA}) + d_i +\sum_{j\neq i } f_{j}(s_{jA}),
\end{equation} for all vectors $\ds s_A= (s_{1A}, s_{2A}, \dots, s_{(i-1)A}, s_{iA}, \dots , s_{nA}) \in\mathbb R^n$ .

Regroup all the coefficients of every term $\ds f_{j}(s_{jA})$ in the relation above, and denote $\ds X_{-i} = (x_{i1}, x_{i2}, \dots x_{i(i-1)}, x_{i(i+1)}, \dots x_{in})$ . 

Also, for every $\ds j\neq i$, let $\ds C_j= \left(
\begin{array}{c}
1\\
\vdots \\
1\\
c_j\\
\vdots\\
1\\
\end{array}
\right)$ be the $\ds 1\times (n-1)$ vector with $\ds j^{\mbox{th}}$ component equal to $\ds c_j$ and all other components equal to $\ds 1$. Then relation $\ds (5.5)$ can be rewritten as 

\begin{equation}
\ds x_{ii} - c_i\cdot f_{i}(s_{iA}) - d_i + \sum_{j\neq i} x_{ij}\cdot \left( d_j + f_{i}(s_{iA}) \right) + \sum_{j\neq i} f_j(s_{jA})\left(X_{-i}\cdot C_{j} - 1\right) =0, 
\end{equation}
for all vectors $\ds s_A= (s_{1A}, s_{2A}, \dots, s_{(i-1)A}, s_{iA}, \dots , s_{nA}) \in\mathbb R^n$ . Note that each term $\ds X_{-i}\cdot C_j$ represents the product of an $\ (n-1)\times 1$ vector with a $\ds 1\times (n-1)$ vector.

Since relation $\ds (5.6)$ holds for all values $\ds s_{1A}, s_{2A}, \dots s_{(i-1)A}, s_{(i+1)A}, \dots, s_{nA}$, and since functions $\ds f_{j}$ are continuous and nonconstant, we must have $\ds X_{-i} \cdot C_j =1$ for all $\ds j\neq i$, and $\ds x_{ii} = c_i\cdot f_{i}(s_{iA}) + d_i - \sum_{j\neq i} x_{ij}\cdot d_j $. These conditions can be rewritten as 

\begin{equation}
 (x_{i1}, x_{i2}, \dots x_{i(i-1)}, x_{i(i+1)}, \dots x_{in}) \cdot  \left( \begin{array}{ccccc}
c_1 & 1 & \dots & 1 &1 \\
1 & c_2 & \dots & 1 & 1 \\
 &  & \vdots& & \\
1 & 1 & \dots & 1 & c_n \end{array} \right) = \left(1, 1, \dots, 1 \right)
\end{equation} and 

\begin{equation}
\ds x_{ii} = f_{i}(s_{iA})\left(c_i -\sum_{j\neq i} x_{ij}\right) + d_i - \sum_{j\neq i} x_{ij}\cdot d_j 
\end{equation}

The matrix in relation $\ds (5.7)$ has elements $\ds c_{j} $, $\ds j\neq i$ in order on the main diagonal, and $\ds 1$s elsewhere. Since the elements $\ds c_i$ are all strictly greater than $\ds 1$, the determinant of the matrix in $\ds (5.7)$ is strictly positive, so relation $\ds (5.7)$ uniquely determines coefficients $\ds x_{ij}$ for $\ds j\neq i$. Further, relation $\ds (5.8)$ now determines $\ds x_{ii}$, so the truthful bidding function is indeed unique, and so the lemma is proved $\ds\square$.

Note that we do not in fact need the functions $\ds f_i$ to be continuous and nonconstant: we merely desire that the set of all values of functions $\ds f_j(s_{jA})$ have full dimension $\ds n-1$ in $\ds \mathbb R^{n-1}$. 

Let us now prove some more interesting properties of the coefficients of truthful bidding functions.

\begin{lemma}
Consider the following system of equations :

$$\ds a_1\cdot x_1 + x_2+ x_3+\cdots+ x_n =1\eqno (1)$$

$$\ds x_1 + a_2\cdot x_2 + x_3+\cdots+ x_n=1\eqno(2)$$

$$\ds x_1+x_2+a_3\cdot x_3+\cdots + x_n =1\eqno(3)$$

$$\ds\vdots$$

$$\ds x_1+x_2+\cdots+ x_{i-1} + a_i\cdot x_i+ x_{i+1}+\cdots+ x_n =1 \eqno(i)$$

$$\ds \vdots$$

$$\ds x_1+ x_2+\cdots+ x_{n-1}+ a_n\cdot x_n =1\eqno (n)$$
\end{lemma}

If all coefficients $\ds a_i$ are strictly greater than $\ds 1$, then each $\ds x_i$ is strictly positive.

\emph{Proof.} First note that by subtracting relations $\ds (i)$ and $\ds (j)$ we obtain 

$\ds (a_i -1)\cdot x_i = (a_j-1)\cdot x_j$, so all the numbers $\ds x_i$ have the same sign, and none of them can be zero, because if at least one of them is zero, then all of them are zero, and the system of equations is no longer satisfied. Also, the numbers $\ds x_i$ cannot be negative, because in that case $\ds a_1\cdot x_1 + x_2+ x_3+\cdots+ x_n <0,$ again contradicting the system of equations.

Therefore the solution $\ds (x_1, x_2, \dots, x_n)$ to this system of equations has all components strictly positive.

Note that if coefficients $\ds (x_{i1}, x_{i2}, \dots x_{in})$ determine the truthful bidding function of buyer $\ds i$ with signal $\ds s_{iA}$, then the $\ds x_{ii}$ component is the only one which can differ as a function of buyer $\ds 1$'s signal $\ds s_{iA}$, while the other components satisfy the property 

\begin{equation}
\ds \frac{1}{x_{ij}}\cdot \left( 1- \sum\limits_{j\neq t\neq i} x_{it} \right) = c_j, \mbox{ for all } j\neq i, j\in N
\end{equation}

We are now ready to describe our efficient auction mechanisms. Again assume that there are at least as many buyers as goods, and split the problem in two parts.

\subsection{ There is an equal number of buyers and goods}

Consider the following auction mechanism (\emph{Auction 3}):

\begin{enumerate}

\item \emph{Bidding.} For every good $\ds A\in M$, every buyer $\ds i\in N$ submits a linear bid function $\ds b_{iA} :\mathbb R^{n-1}\mapsto \mathbb R,$ such that for any $\ds v_{-i} = (v_1, v_2, \dots, v_{i-1}, v_{i+1}, \dots, v_{n})\in \mathbb R^{n-1}$, we have
 
\begin{equation}
\ds b_{iA} (v_{-i}) = x_{ii}^{(A)} + \sum_{j\neq i} x_{ij}^{(A)}\cdot v_j, 
\end{equation}
 for some coefficients $\ds x_{i1}^{(A)}, x_{i2}^{(A)}, \dots x_{in}^{(A)}\in\mathbb R$

\item \emph{Allocation.} The goods in $\ds M$ are allocated if and only for every good $\ds A\in M$, $\ds x_{ij}^{(A)}\neq 0$ for all $\ds i\neq j$, and for every buyer $\ds j\in N$, there exists a value $\ds c_j'>1$ such that

\begin{equation}
\ds \frac{1}{x_{ij}^{(A)}}\cdot \left( 1- \sum\limits_{j\neq t\neq i} x_{it}^{(A)} \right) =  \frac{1}{x_{kj}^{(A)}}\cdot \left( 1- \sum\limits_{j\neq t\neq k} x_{kt}^{(A)} \right) = c_j', \mbox{ for all } A\in M, i, k\in N \mbox{ and } i,k\neq j
\end{equation}

For each good $\ds A\in M$, consider the fixed point $\ds v_A^{\circ} = (v_{1A}^{\circ}, v_{2A}^{\circ}, \dots, v_{nA}^{\circ}),$ such that for all $\ds i\in N$:

\begin{equation}
b_{iA}(v_{1A}^{\circ}, v_{2A}^{\circ}, \dots, v_{(i-1)A}^{\circ}, v_{(i+1)A}^{\circ}, \dots v_{nA}^{\circ}) = v_{iA}^{\circ}
\end{equation}

Each allocation of the goods to the buyers corresponds to a permutation $\ds \sigma \in P_n$ such that buyer $\ds i$ receives good $\ds \sigma(i)$.
Under the assumption that $\ds v_{iA}^{\circ}$ represents buyer $\ds i$'s valuation of good $\ds A$, let $\ds \sigma^*$ be the welfare-maximizing allocation of the goods in $\ds M$ to the buyers in $\ds N$. Then buyer $\ds i$ receives good $\ds \sigma^*(i)$. 

\item \emph{Payment.} Define $\ds W(\sigma) = \sum_{j\in N} v_{j\sigma(j)}^{\circ}$, the apparent welfare from allocation $\ds \sigma$. Finally, let 

\begin{equation}
P_i(\sigma) = \frac{c_i'}{c_i'-1} \cdot W(\sigma) - v_{i\sigma(i)}^{\circ} + \frac{c_i'}{c_i' -1}\cdot\frac{1}{c_{i}' -\sum\limits_{j\neq i}x_{ij} }\cdot \sum_{A\in M} x_{ii}^{(A)}
\end{equation}

If $\ds \sigma^*$ is the welfare-maximizing allocation selected in step $\ds 2$, buyer $\ds i$ pays 

\begin{equation}
\ds \Max_{\sigma\in P_n} P_i(\sigma) - P_i(\sigma^*)
\end{equation}

\end{enumerate}
Let's prove that this auction is well-defined. First of all, note that condition $\ds (5.11)$ is equivalent to saying that condition $\ds (5.9)$ holds for every buyer $\ds k\neq j$. This forces buyers to bid as if they believe that $\ds \frac{\partial w_i}{\partial f_i} = c_i'$ for all $\ds i\in N$. Therefore, bid functions satisfying relation $\ds (5.11)$ correspond to truthful bidding functions in the case  $\ds \frac{\partial w_i}{\partial f_i} = c_i'$, while the free terms $\ds x_{ii}^{(A)}$ correspond to each buyer's value of $\ds f_{i}(s_{iA})$, from relation $\ds (5.8)$. Let us prove another useful lemma:

\begin{lemma}
If bid functions respect condition $\ds (5.11)$, then there is a unique fixed point satisfying $\ds (5.12)$ .

\end{lemma}

\emph{Proof.}

Let $\ds (v_1, v_2, \dots v_n)$ be a fixed point satisfying $\ds (5.12)$ for a given good $\ds A$, and write $\ds x_{ij}$ instead of $\ds x_{ij}^{(A)}$ for simplicity. $\ds (5.12)$ This is equivalent to 

\begin{equation} \left( \begin{array}{ccccc}
-1 & x_{12} & x_{13} & \dots &x_{1n} \\
x_{21} & -1 & x_{23} & \dots & x_{2n} \\
 &  & \vdots& & \\
x_{n1} & x_{n2} & x_{n3} & \dots & -1 \end{array} \right) \cdot  \left( \begin{array}{c}
v_1 \\
v_2 \\
\vdots \\
v_n\\\end{array} \right) =  \left( \begin{array}{c}
-x_{11}\\
-x_{22} \\
\vdots \\
-x_{nn}\\\end{array} \right)
\end{equation}

However, note that from $\ds (5.11)$ we obtain
$$\ds
 \left( \begin{array}{ccccc}
-1 & x_{12} & x_{13} & \dots &x_{1n} \\
x_{21} & -1 & x_{23} & \dots & x_{2n} \\
 &  & \vdots& & \\
x_{n1} & x_{n2} & x_{n3} & \dots & -1 \end{array} \right) \cdot \left( \begin{array}{ccccc}
c_1' & 1 & 1 & \dots & 1 \\
1 & c_2' & 1 & \dots & 1 \\
 &  & \vdots& & \\
1 & 1 & 1 & \dots & c_n' \end{array}\right) = $$

\begin{equation} =\left( \begin{array}{ccccc}
-c_1' +\sum\limits_{j\neq 1} x_{1j} & 0 & 0 & \dots & 0 \\
0 & -c_2'+\sum\limits_{j\neq 2} x_{2j} & 0 & \dots & 0 \\
 &  & \vdots& & \\
0 & 0 & 0 & \dots & -c_n' +\sum\limits_{j\neq n} x_{nj} \end{array}\right),
\end{equation}

The latter matrix is zero everywhere except on the main diagonal, where entry $\ds i$ equals $\ds -c_i' +\sum_{j\neq i} x_{ij}$.
Using lemma $\ds (5.2)$ for coefficients $\ds c_i'$ in $\ds (5.11)$, we obtain that $\ds -c_i' +\sum_{j\neq i} x_{ij} <0 $ for all $\ds i$. Therefore the matrix is invertible, and hence the matrix in relation $\ds (5.15)$ is invertible, so there always exists a fixed point given condition $\ds (5.11)$, and that fixed point is unique $\ds\square$.

Note that since $\ds c_i' > \sum_{j\neq i} x_{ij}$, equation $\ds (5.13)$ is also well-defined.

Finally, note that given values $\ds x_{i1}, x_{i2}, \dots x_{in}, d_1, d_2, \dots d_n, c_i'$, there is a unique $\ds x\in \mathbb R$ with $\ds  x = f_i(s_{iA'})$, for some $\ds s_{iA}'\in\mathbb R$,  such that

\begin{equation}
\ds x_{ii} = f_{i}(s_{iA})\left(c_i' -\sum_{j\neq i} x_{ij}\right) + d_i - \sum_{j\neq i} x_{ij}\cdot d_j 
\end{equation}

This means that equation $\ds (5.8)$ completely determines $\ds f_i(s_{iA})$, and if the function $\ds f_i$ is injective, this uniquely determines signal $\ds s_{iA}'$ for which equation $\ds (5.8)$ holds. If additionally $\ds Im(f_i) =\mathbb R$, then any value of $\ds x_ii$ corresponds to a different signal value $\ds s_{iA}'$ for which $\ds (5.17)$ holds.

We are now ready to prove the main result of this subsection.

\begin{theorem} Assume $\ds n\geq 3$.
If valuation functions satisfy conditions $\ds (4.1) - (4.5)$, $\ds (5.2)$ and if each function $\ds f_i$ is injective and $\ds Im(f_i)=\mathbb R$, then truthful bidding is a Nash Equilibrium of Auction 3, and the auction is efficient.
\end{theorem}

\emph{Proof.} Assume that each buyer $\ds j$ has signal $\ds s_{jA}$ for a given good $\ds A\in M$, and assume that each buyer $\ds j\neq i$ bids truthfully, i.e. bids according to $\ds (5.8)$ and $\ds (5.9)$. Then condition $\ds (5.11)$ is satisfied for buyers $\ds j\neq i$, and the constants are $\ds c_i$. 

We distinguish two simple cases:

1. buyer $\ds i $ bids linear functions such that at least one of them does not satisfy $\ds (5.11)$. In this case no good is allocated, and so buyer $\ds i$ receives zero utility.

2. buyer $\ds i$ bids linear functions satisfying $\ds (5.11) $. Hence, the only freedom that buyer $\ds i$ has is in choosing coefficients $\ds x_{ii}^{(A)}$, and since each coefficient $\ds x_{ii}^{(A)}$ corresponds to a different signal value $\ds s_{iA}'$, buyer $\ds i$ is essentially bidding truthfully as if his or her signal were $\ds s_{iA}'$, and every other buyer $\ds j\neq i$ bids truthfully his or her signal $\ds s_{jA}$ for every $\ds A\in M$. Therefore the fixed points simply represent valuation functions for each buyer and each good $\ds A$ as if buyer $\ds i$ has signal $\ds s_{iA}'$ and every other buyer $\ds j$ has signal $\ds s_{jA}$.

In this case, the allocation rule in this auction is equivalent to the allocation rule in \emph{Auction 1} in section 4, while the payment rule is also equivalent to the payment rule in \emph{Auction 1} . Note that definition $\ds (5.13)$ of $\ds P_{\sigma}$ is the same as condition $\ds (4.7)$ except for a constant term $\ds m\cdot \frac{c_i}{c_i-1}\cdot\frac{1}{c_i -\sum\limits_{j\neq i} x_{ij}} \left( d_i - \sum_{j\neq i}x_{ij}\cdot d_j\right)$, which gets cancelled out when considering the final payment $\ds (5.17)$. 

Therefore, in this second case, our \emph{Auction 3} is equivalent to \emph{Auction 1}, where every buyer receives nonnegative utility, so there is no incentive to deviate to case 1 above, and truthful bidding represents is a best response of buyer $\ds i$, as desired.

Therefore truthful bidding is indeed a Nash Equilibrium of this auction, and the auction is efficient $\ds \square$.

Note that if $\ds n=2$, condition $\ds (5.11)$ does not uniquely determine $\ds x_{12}$ for buyer $\ds 1$, since there is only one buyer for which the relation holds. Thus the reasoning above does not hold. Nevertheless, there is an efficient auction mechanism in this case : keep the bidding and allocation rules the same, but define the payment in the following way:

Assume the buyers are $\ds 1, 2$ and the goods $\ds A, B$ . Let $\ds\sigma_1$ represent allocation $\ds (A, B)$ and $\ds\sigma_2$ represent allocation $\ds (B, A)$, with $\ds \sigma_1(1) =\sigma_2(2) = A, \sigma_1(2)=\sigma_2(1) =B$. For allocation $\ds\sigma \in P_2$, where buyer $\ds i$ receives good $\ds \sigma(i)$, define 

\begin{equation}
\ds P_i(\sigma) = \frac{c_i'}{c_i'-1}\cdot x_{jj}^{\sigma(j)}
\end{equation}

If allocation $\ds\sigma^*$ is selected, buyer $\ds i$ pays 
\begin{equation} max \left (P_i(\sigma_1), P_i(\sigma_2)\right) - P_i(\sigma^*)
\end{equation}

Truthful bidding is a Nash Equilibrium of this allocation because, if buyer $\ds 2$ plays truthfully, then $\ds x_{22}^{A} - x_{22}^{B} = \frac{c_1\cdot c_2-1}{c_1}\left(f_2(s_{2A})- f_2(s_{2B})\right)$, and 

$\ds P_1(\sigma_1)- P_1(\sigma_2) = \frac{c_1\cdot c_2 -1}{c_1-1}\left(f_2(s_{2B})- f_2(s_{2A})\right)$. 

Thus buyer $\ds 1$ has an incentive to bid so that allocation $\ds \sigma_1$ is selected if and only if $$\ds c_1\cdot f_1(s_{1A}) + f_2(s_{2A}) + P_1(\sigma_1) \geq c_1\cdot f_1(s_{1B}) + f_2(s_{2B}) + P_1(\sigma_2)$$ $$\ds \iff c_1( f_1(s_{1A})- f_1(s_{1B})) \geq \frac{c_1(c_2-1)}{c_1-1},$$ which holds if and only if the overall welfare from allocation $\ds\sigma_1$ is higher than the overall welfare from allocation $\ds \sigma_2$.

Thus truthful bidding is a best response for buyer $\ds 1$ if buyer $\ds 2$ truthfully, and since the argument is symmetric, truthful bidding does indeed constitute a Nash Equilibrium.

Let us now deal with the case of more buyers than goods.

\subsection{There are more buyers than goods}

Assume there is a set $\ds N$ of $\ds n$ buyers and a set $\ds M$ of $\ds m$ goods, with $\ds n>m$.

\bigskip

Consider the following auction mechanism (\emph{Auction 4}):

\begin{enumerate}

\item \emph{Bidding.} For every good $\ds A\in M$, every buyer $\ds i\in N$ submits a linear bid function $\ds b_{iA} :\mathbb R^{n-1}\mapsto \mathbb R,$ such that for any $\ds v_{-i} = (v_1, v_2, \dots, v_{i-1}, v_{i+1}, \dots, v_{n})\in \mathbb R^{n-1}$, we have
 
\begin{equation}
\ds b_{iA} (v_{-i}) = x_{ii}^{(A)} + \sum_{j\neq i} x_{ij}^{(A)}\cdot v_j, 
\end{equation}
 for some coefficients $\ds x_{i1}^{(A)}, x_{i2}^{(A)}, \dots x_{in}^{(A)}\in\mathbb R$

\item \emph{Allocation.} The goods in $\ds M$ are allocated if and only for every good $\ds A\in M$, $\ds x_{ij}^{(A)}\neq 0$ for all $\ds i\neq j$, and for every buyer $\ds j\in N$, there exists a value $\ds c_j'>1$ such that

\begin{equation}
\ds \frac{1}{x_{ij}^{(A)}}\cdot \left( 1- \sum\limits_{j\neq t\neq i} x_{it}^{(A)} \right) =  \frac{1}{x_{kj}^{(A)}}\cdot \left( 1- \sum\limits_{j\neq t\neq k} x_{kt}^{(A)} \right) = c_j', \mbox{ for all } A\in M, i, k\in N \mbox{ and } i,k\neq j
\end{equation}

For each good $\ds A\in M$, consider the fixed point $\ds v_A^{\circ} = (v_{1A}^{\circ}, v_{2A}^{\circ}, \dots, v_{nA}^{\circ}),$ such that for all $\ds i\in N$:

\begin{equation}
b_{iA}(v_{1A}^{\circ}, v_{2A}^{\circ}, \dots, v_{(i-1)A}^{\circ}, v_{(i+1)A}^{\circ}, \dots v_{nA}^{\circ}) = v_{iA}^{\circ} .
\end{equation}

Under the assumtion that the $\ds i^{\mbox{th}}$ component of fixed point vector $\ds v_A^{\circ}$ represents each buyer $\ds i$'s valuation of good $\ds A$, let $\ds S= (S_1, S_2, \dots S_n)$ be the the welfare-maximizing allocation, such that any good $\ds A\in M$ is offered to buyer $\ds i_A\in N$, and each buyer receives at most one good. Then goods are offered according to allocation $\ds S$.

\item \emph{Payment.} If buyer $\ds i$ was assigned no goods at step $\ds 2$ above, he or she pays nothing. If buyer $\ds i$ was assigned good $\ds A$ based on step (2) above, let $\ds S'_{-i} = (S_1', S_2', \dots ,S_n')$ be the apparent welfare-maximizing allocation of the goods in $\ds M$ to the buyers $\ds N - \{i\}$, according to the same assumptions as in step $\ds 2$, such that every good $\ds K$ is assigned to buyer $\ds j_K$. 

Remember that $\ds S$ is the welfare-maximizing allocation according to the fixed points in $\ds (2.2)$, and assume buyer $\ds i$ receives good $\ds A$, and every good $\ds K\neq A$ is offered to buyer $\ds i_K$.

Consider the bidding $\ds x_{ii}^{A*}$ which would induce the fixed point  $\ds v_A^{*} = (v_{1A}^{*}, v_{2A}^{*}, \dots, v_{nA}^{*})$ satisfying the fixed point condition $\ds (5.22)$ and 

\begin{equation}
v_{iA}^* +\sum_{K\neq A} v_{i_KK}^{\circ} = v_{j_AA}^* +\sum_{K\neq A} v_{j_KK}^{\circ}
\end{equation}

Then buyer $\ds i$ pays $\ds v_{iA}^*$.

\end{enumerate}

Based on our work on the previous auction, it is easy to prove that this auction is efficient.

\begin{theorem}
If valuation functions satisfy conditions $\ds (4.1) - (4.5)$, $\ds (5.2)$ and if each function $\ds f_i$ is injective and $\ds Im(f_i)=\mathbb R$, then truthful bidding is a Nash Equilibrium of Auction 4, and the auction is efficient.
\end{theorem}

\emph{Proof.}

We simply note that again, if $\ds n\geq 3$ and all buyers but $\ds i$ bid truthfully, then buyer $\ds i$ is essentially forced to submit a bidding function as if his signal was the value $\ds s_{iA}$ satisfying $\ds (5.17)$. The allocation and payment components of this auction are equivalent to the allocation and payment components of auction $\ds 2$, therefore if every buyer $\ds j\neq i$ bids truthfully, then this auction is equivalent to auction 2, so buyer $\ds i$'s best response is to bid truthfully. Therefore truthful bidding is a Nash Equilibrium and this auction is efficient.

Let's prove this result also in case $\ds n=2$, although this case was proven in $\ds [6]$ under more general conditions.

Since there is only one good, in case buyer $\ds 1$ wins, he or she makes a payment of $\ds v_2^*$, where $\ds v_2^* = b_2(v_2^*)$.

Assume buyer $\ds 2$ bids truthfully. Then we have $$\ds v_1(s_{1A}, s_{2A}) \geq v_2^* \iff v_{1}(s_{1A}, s_{2A}) - v_2^* \geq \frac{1}{c_1}\cdot\left( v_{1}(s_{1A}, s_{2A}) - v_2^* \right)$$ $$\ds\iff v_{1}(s_{1A}, s_{2A}) - v_2^* \geq b_2(v_1(s_{1A}, s_{2A})) - b_2(v_2^*) $$ $$\ds \iff  v_1(s_{1A}, s_{2A}) \geq v_2(s_{1A}, s_{2A}).$$

Therefore buyer $\ds 1$ receives utility from winning the good if and only if his actual valuation for the good is greater than buyer $\ds 2$'s, so indeed truthful bidding is a best response, and so truthfull bidding represents a Nash Equilibrium $\ds\square$.

\section{Conclusion}

In this paper we have provided efficient auction mechanisms in the case of linear, separable valuation functions, under the assumption of scarcity ( as least as many buyers than goods) and under the assumption that no buyer gains from having more than one good. We also showed that except for linearity and separability, the other assumptions are necessary for an efficient mechanism to exist. It would be very interesting to try to extend the mechanisms in this paper in the case where valuations are nonlinear and also in the case of more goods than buyers, even though the latter is not usually true in practice.

I am very grateful to Professor Barry Mazur for very useful comments and to Professor Eric Maskin for all his patience and guidance throughout this entire process.

\end{document}